\documentclass[aps,prf,onecolumn,10pt,superscriptaddress]{revtex4-2}
\usepackage{amsmath}
\usepackage{graphicx}
\usepackage{color}
\usepackage{hyperref}
\usepackage{appendix}

\usepackage{geometry}
\begin{document}

\title{Effect of a weak current on wind-generated waves in the wrinkle regime}
\author{C. Nov\'e-Josserand}
\affiliation{Universit\'e Paris-Saclay, CNRS, FAST, 91405, Orsay, France.}
\affiliation{LadHyX, UMR CNRS 7646, Ecole Polytechnique, 91128 Palaiseau, France}
\author{S. Perrard}
\affiliation{LPENS, D\'epartement de Physique, Ecole Normale Sup\'erieure, PSL University, 75005 Paris France}
\author{A. Lozano-Dur\'an}
\affiliation{Center for Turbulence Research, Stanford University, Stanford, California 94305, USA}
\author{M. Benzaquen}
\affiliation{LadHyX, UMR CNRS 7646, Ecole Polytechnique, 91128 Palaiseau, France}
\author{M. Rabaud}
\affiliation{Universit\'e Paris-Saclay, CNRS, FAST, 91405, Orsay, France.}
\author{F. Moisy}
\email{frederic.moisy@universite-paris-saclay.fr}
\affiliation{Universit\'e Paris-Saclay, CNRS, FAST, 91405, Orsay, France.}

\date{\today}

\begin{abstract}
\vskip 6mm

We investigate numerically the influence of a weak current on wind-generated surface deformations for wind velocity below the onset of regular waves. In that regime, the liquid surface is populated by small disorganised deformations elongated in the wind direction, referred to as \textit{wrinkles}. These wrinkles are the superposition of incoherent wakes generated by the pressure fluctuations traveling in the turbulent boundary layer in the air. In this work, we account for the effect of a weak sheared current in the liquid, either longitudinal or transverse, by introducing a modified Doppler-shifted dispersion relation to lowest order in viscosity and current in the spectral theory previously derived by Perrard~{\it et al.} [J. Fluid Mech. {\bf 873}, 1020 (2019)]. This theory describes the simplified one-way problem of surface deformations excited by a prescribed turbulent forcing, thereby neglecting the retroaction of waves on turbulence in the air. The forcing is taken from a set of direct numerical simulations of a turbulent channel flow. We determine the wrinkle properties (size and amplitude) as a function of the liquid viscosity and current properties (surface velocity, thickness and orientation). We find significant modifications of the wrinkle geometry by the currents: the wrinkles are tilted for a transverse current, and show finer scales for a longitudinal current. However, their characteristic size is weakly affected, and their amplitude remains independent of the current. We discuss the implications of these results on the onset of regular waves at larger wind velocity.  In this work, we introduce a spectral interpolation method to evaluate the surface deformation fields, based on a refined meshing close to the dispersion relation of the waves. This method, which can be extended to any dispersive system excited by a random forcing, strongly reduces the discretization effects at a low computational cost.  

\end{abstract}

\maketitle
	
\section{Introduction}
\label{sec:intro}

When a light turbulent wind blows at the surface of a liquid at rest, it first generates random surface deformations of weak amplitude elongated in the wind direction~\cite{Russell_1844,Keulegan_1951,Phillips_1957,Gottifredi_1970,Kahma_1988,Zhang_1995,Banner_1998,Caulliez_2008}. These structures, named {\it wrinkles} in Refs.~\cite{Paquier_2015, Paquier_2016},  can be described as the superposition of the incoherent wakes originating from the pressure and shear stress fluctuations traveling in the turbulent boundary layer in the air~\cite{Perrard2019}. If the wind is sufficiently strong, typically 1-3~m~s$^{-1}$ for the air-water interface, these wrinkles are found at small fetch only, and rapidly evolve downwind into more coherent waves of larger amplitude. On the other hand, if the wind velocity remains low, these wrinkles reach a statistically stationary state, in which the energy injected by the pressure fluctuations that push or suck the surface is balanced by the energy dissipated in the liquid. This statistically steady state corresponds to the asymptotic regime of the inviscid resonant theory of Phillips~\cite{Phillips_1957} saturated by the viscous dissipation in the liquid.

Although these incoherent surface deformations at small wind velocities have been observed for a long time, their very small amplitude (typically $1-10~\mu$m in water), well below the resolution of conventional probes, make them difficult to analyze experimentally. Wrinkles are also found in numerical simulations of temporally growing waves, but the  range  of  physical  parameters  covered  by  these  studies remains  limited~\cite{Lin_2008,Zonta_2015}.
They were systematically characterized by Paquier {\it et al.}~\cite{Paquier_2015,Paquier_2016} in water and more viscous aqueous solutions using free-surface synthetic Schlieren measurements~\cite{Moisy_2009}, an optical method with micrometer accuracy.

The motivation for investigating wind-wave generation in the wrinkle regime is that, despite their very small amplitude, wrinkles may play a key role in the onset of coherent regular  waves at larger wind velocity. If wrinkles are the base state from which regular waves grow as the wind velocity is increased, we may expect the transition to regular waves to depend on any parameter that may affect the wrinkles, such as the presence of currents in the liquid.

Beyond their relevance for oceanography, wrinkles are also of interest for industrial applications that involves thin liquid films sheared by turbulent gas flow, such as coating processes, cooling of solidifying surfaces, and two-phase flows in oil industry~\cite{Fulgosi2003,Vellingiri2013}.  Although the wave dynamics in thin films strongly differs from that in the deep-water limit relevant to the air-sea configuration, elongated wrinkles produced by the wakes of pressure and stress fluctuations are also observed in that configuration~\cite{Bender2019}.

The theoretical and numerical analysis of Perrard {\it et al.}~\cite{Perrard2019} identified the main scaling properties of the wrinkles in deep water in the absence of currents. Their characteristic size $\Lambda$ is governed by the largest scales of the pressure fluctuations, controlled by the thickness $\delta$ of the boundary layer, with no significant effect of the liquid viscosity $\nu_\ell$. On the other hand, their characteristic amplitude $\zeta_{\mathrm{rms}} = \langle \zeta^2 \rangle^{1/2}$ [with $\zeta({\bf r},t)$ the surface displacement field] depends on $\nu_{\ell}$: in the statistically steady state, the balance between the work of the pressure fluctuations per unit time and the dissipation in the liquid yields
\begin{equation}
\frac{\zeta_{\mathrm{rms}}} {\delta} \simeq C \frac{\rho_a}{\rho_\ell} \left( \frac{{u^*}^{3}}{g \nu_\ell} \right)^{1/2},
\label{eq:wa}
\end{equation} 
with $C \simeq 0.02$~\cite{Perrard2019}. Here $u^*$ is the friction velocity in the air (one has $u^* \simeq 0.05 U_a$ for the typical Reynolds number of the problem, with $U_a$ the freestream velocity), $g$ the acceleration of gravity, and $\rho_a$ and $\rho_\ell$ the density of air and liquid; the liquid depth is assumed infinite, and the capillary effects are neglected, provided that the boundary layer thickness $\delta$ is much larger than the capillary length.

Equation~(\ref{eq:wa}) is in good agreement with laboratory experiments over a wide range of liquid viscosity, $\nu_\ell = 1-560$~mm$^2$s$^{-1}$~\cite{Paquier_2016}. However, extending laboratory results, for which the boundary layer thickness is typically $\delta \simeq 1-10$~cm,  to the ocean is challenging, because of the difficulty  to evaluate the spatio-temporal structure of pressure fluctuations in the atmospheric boundary layer.  The thickness of the boundary layer over the ocean is usually governed by unsteady conditions or convection phenomena~\citep{kaimal1976turbulence}; values of order $100 - 500$~m reported in the literature~\cite{peng2016detecting,stull2012introduction} are order of magnitudes larger than the centimetric surface deformations typically observed.

An important limitation of the theory in Ref.~\cite{Perrard2019} is that it ignores the presence of currents in the liquid: only the stress fluctuations (pressure and shear stress) are considered, while the mean shear stress applied by the wind, responsible for the generation of a surface current, is neglected~\cite{sullivan2010dynamics}.  Stationary currents in the liquid, not necessarily aligned with the wind, are frequently encountered in natural flows, such as in near-shore regions and rivers~\cite{peregrine1976interaction,dong2012theoretical,ellingsen2016oblique}.  In the case of wind-generated drift flow, the surface velocity $U_s$ results from a balance between the applied wind stress and the viscous stress in the fluid (Stokes-drift contribution is usually negligible in that context~\cite{wu1975wind,rascle2008global}). Wind-generated currents are typically of order $0.6 u^*$~\cite{wu1983sea,veron2001experiments,caulliez2007turbulence,shemer2019evolution}, but currents originating from other external causes may naturally be significantly larger than $u^*$.

Modeling the combined effects of the {\it mean} shear stress, responsible for the generation of a current, and the {\it fluctuating} stresses (including wave-induced stresses) is of considerable difficulty in air-sea interaction~\cite{Longuet1969,Banner_1998}.
In this paper we consider a simplified configuration, valid only in the wrinkle regime, following the assumptions introduced in Perrard {\it et al.}~\cite{Perrard2019}: (1) we neglect the feedback of the waves on the turbulent boundary layer (one-way approach), an assumption valid when the wrinkle amplitude is much smaller than the viscous sublayer thickness; (2) we assume that the flow in the liquid is laminar, which allows us to consider separately the effect of the sheared current and the waves; (3) we neglect the effect of the shear stress fluctuations, which were found to produce surface deformations much smaller than that produced by the pressure fluctuations. In this simplified configuration, the sheared current is simply modeled through a modification of the dispersion relation of the waves. Considering separately the mean sheared current and the surface deformations induced by the stress fluctuations is valid only for sufficient viscosity, as in the experiments of Paquier {\it et al.}~\cite{Paquier_2015,Paquier_2016} performed in viscous aqueous solutions. It is however questionable in the real air-sea interaction problem, in which even a moderate wind produces a highly sheared and possibly turbulent layer at the surface of the water.

In this paper, we are interested in the modifications of the wrinkle amplitude and geometry induced by such a shear-modified dispersion relation. Since wrinkles are elongated in the wind direction,  we can anticipate a stronger influence of a crosswind current than an alongwind current: the dominant wave number ${\bf k}$ of the wrinkles being approximately normal to the wind direction, a stronger Doppler shift ${\bf k} \cdot {\bf U}_s$ is naturally expected for a current ${\bf U}_s$ normal to the wind.

Several approaches, all assuming linear inviscid waves, were introduced to determine the modification of the dispersion relation owing to sheared currents. Solutions to this problem are either analytical or numerical~\cite{LiEllingsen2019}. Analytical approaches are based on a perturbation analysis for weak currents, valid to first or second order in $U_s/c$ (with $c$ the phase velocity)~\cite{stewart1974hf,skop1987approximate,kirby1989surface,shrira1993surface}. Numerical schemes include piecewise linear approximation for the velocity profile~\cite{zhang2005short, smeltzer2017surface}, or a full Rayleigh approach for arbitrary velocity profile~\cite{dong2012theoretical}. Recently, Li and Ellingsen~\cite{LiEllingsen2019} introduced a theoretical and numerical method that works for arbitrary velocity profiles including slowly varying bathymetry.

We restrict in this paper to the effect a weak sheared current on the wrinkle properties. The influence of a sheared current on the surface deformation induced by a traveling pressure disturbance is analyzed in Ref.~\cite{LiSmeltzer2017}, but without viscous effects. Viscosity must naturally be kept in our analysis, since wrinkles are the viscous-saturated statistically steady state of waves sustained by the turbulent fluctuations in the air. We propose here a heuristic modification of the spectral theory of Perrard~{\it et al.} \cite{Perrard2019} including the effects of viscosity and shear currents.  To provide physical insight, we focus on weak currents, for which the approximate dispersion relation is known analytically to first order in $U_s/c$. Another reason for this restriction is that no exact wave-current interaction analysis including viscous effects is available, so we must consider the problem to lowest order both in viscosity and current. For this reason, we consider in this paper the simplest first-order shear-modified dispersion relation derived by Stewart and Joy~\cite{stewart1974hf}.

In the following we first focus on a uniform current, for which the effect is strongest, and then investigate the more relevant case of a current exponentially decreasing with depth, as sketched in Fig.~\ref{fig:sketch}. We restrict to currents of uniform direction, ignoring the more complex situation of a depth-varying current direction. Our results show that, while the geometry of the wrinkles is modified by currents, their amplitude remains almost independent of the current, suggesting that the wrinkle properties are robust with respect to currents.

In this paper we also introduce  a numerically efficient interpolation method to compute the wrinkle properties. A limitation of the spectral theory in Ref.~\cite{Perrard2019} is that, in deriving Eq.~(\ref{eq:wa}), the limit of small viscosity is taken. This assumption was necessary to derive analytically  the scaling of the wrinkle properties with the liquid viscosity, $\zeta_\mathrm{rms} \sim \nu_\ell^{-1/2}$ and $\Lambda \sim \nu_\ell^0$. This semi-analytical procedure also circumvented the discretisation errors that arise when computing the surface deformation spectrum from direction numerical simulation (DNS) data in boxes of limited size. Such discretisation errors are unavoidable at small $\nu_\ell$, when the resonance is thinner than the spectral resolution of the data. A general procedure was missing to apply this spectral theory to arbitrary viscosity, or more generally to arbitrary dispersive wave system for which partial analytical solutions cannot be derived. Here we propose an improved version for the evaluation of the surface deformation spectrum which does not assume weak viscosity, based on an interpolation of the forcing spectrum in the vicinity of the resonance. Using this method, the dependence of the wrinkle properties in liquid viscosity can be investigated, confirming the robustness of the scalings  $\zeta_\mathrm{rms} \sim \nu_\ell^{-1/2}$ and $\Lambda \sim \nu_\ell^0$ derived analytically for small viscosity. This spectral interpolation method could be applied in principle to any physical system governed by dispersive waves excited by a statistically stationnary and homogeneous forcing.

\section{Theoretical description of wrinkles} 
\label{sec:theo}

\subsection{Flow configuration and dimensionless numbers}
\label{sec:ndn}

\begin{figure}[bt]
	\begin{center}
		\includegraphics[width=0.85\linewidth]{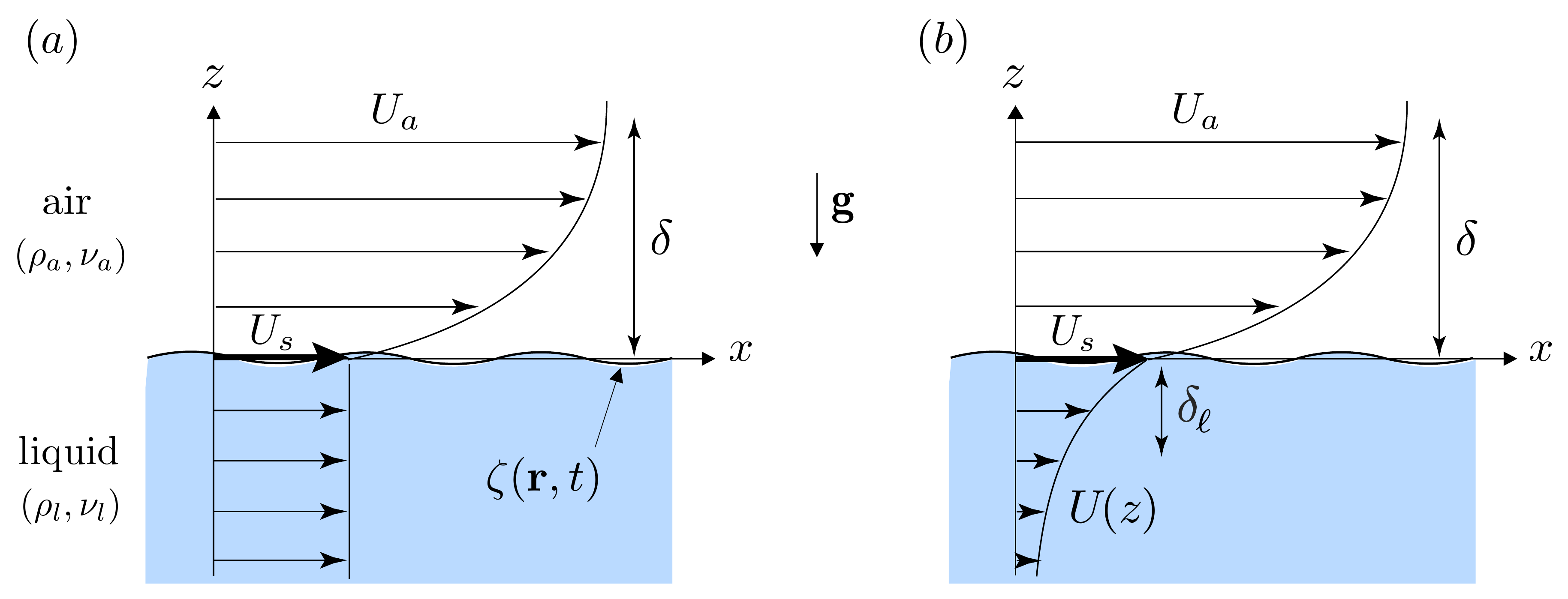}
		\caption{Flow configuration. A liquid is subject to a turbulent air flow blowing at its surface with velocity $U_a$. The liquid is characterized by its density $\rho_\ell$, viscosity $\nu_\ell$, and surface tension $\gamma$. The air turbulent boundary layer is characterized by its density $\rho_a$, kinematic viscosity $\nu_a$, boundary layer thickness $\delta$ and friction velocity $u^*$. Fluctuations in the surface elevation $\zeta({\bf r},t)$ result from  the turbulent stresses applied at the interface. Two current configurations are illustrated: (a) uniform current of constant velocity $U_s$ and (b) exponential velocity profile of thickness $\delta_\ell$. Both cases are illustrated here in the case $\theta=0$ (current aligned with the wind direction). The velocity profiles are not drawn to scale: the velocity in the air is typically 20 times larger than that in the liquid.}
		\label{fig:sketch}
	\end{center}
\end{figure}

We briefly recall here the spectral formulation derived in Ref.~\cite{Perrard2019} that relates the spatio-temporal spectrum of the surface deformation to that of the turbulent forcing.  We first neglect the surface current.

The system is sketched in Fig.~\ref{fig:sketch} with $U_s=0$: a layer of liquid with density $\rho_\ell$, surface tension $\gamma$ and viscosity $\nu_\ell$ is subject to a turbulent wind in the $x$-direction, of density $\rho_a$ and viscosity $\nu_a$. The wind velocity far from the surface is $U_a$, and forms a boundary layer of thickness $\delta$, which we assume to be uniform and statistically stationary (more precisely, we restrict our analysis to length scales and time scales over which $\delta$ can be considered as constant).   The wind applies a shear stress at the surface, of average $\tau_a = \rho_a u^{*2}$, where $u^*$ is the friction velocity.  We neglect for the moment the drift induced by this average shear stress, and focus on the fluctuating stresses at the surface: pressure $p(x,y,z=0,t)$ and shear stress $\boldsymbol{\sigma}(x,y,z=0,t) = \rho_a \nu_a \partial_z {\bf u}_\parallel |_{z=0}$ (where ${\bf u}_\parallel$ is the horizontal velocity fluctuation), with $\langle p \rangle = 0$ and $\langle \boldsymbol{\sigma} \rangle = {\bf 0}$.

The problem without current is characterized by five dimensionless numbers: the density ratio $\rho_a/\rho_\ell$, the Reynolds number  $Re_\delta = u^* \delta / \nu_a$, the Bond number $Bo_\delta = \delta / \ell_c$ (with $\ell_c = \sqrt{\gamma / \rho_\ell g}$ the capillary length), the Froude number $Fr_\delta = u^* / \sqrt{g \delta}$,  and the dimensionless liquid viscosity $\tilde{\nu}_\ell = \nu_\ell / \sqrt{g \delta^3}$. The Froude number characterizes the geometry of wakes generated by the disturbances of size $\delta$ traveling at a characteristic velocity $u^*$: wakes form characteristic V-shaped patterns at small $Fr_\delta$, which narrow at larger $Fr_\delta$~\cite{Rabaud_2013,Darmon_2014}. The normalized liquid viscosity $\tilde{\nu}_\ell$ compares the viscous time scale $\delta^2 / \nu_\ell$ to the period $\sqrt{\delta/g}$ of the gravity wave of wavelength of the order of $\delta$.  We restrict our analysis here to $\tilde{\nu}_\ell \ll 1$, corresponding to weakly damped waves; note that although the viscous effects are weak in the dispersion relation, they are nonetheless essential in the problem, as they govern the saturated wrinkle amplitude. Using this set of dimensionless numbers, the wrinkle amplitude (\ref{eq:wa}) reads
\begin{equation}
\frac{\zeta_{\mathrm{rms}}} {\delta} \simeq C \frac{\rho_a}{\rho_\ell} \tilde{\nu}_\ell^{-1/2} Fr_\delta^{3/2}.
\label{eq:wab}
\end{equation}

In air-water laboratory experiments and in the ocean, we have $\rho_a/\rho_\ell \simeq 1.2 \times 10^{-3}$, $Re_\delta \gg 1$, $Bo_\delta \gg 1$, $Fr_\delta \simeq O(1)$, and $\tilde{\nu}_\ell \ll 1$.  If we choose $\delta = 3$~cm as in the experiments of Paquier {\it et al.}~\cite{Paquier_2015,Paquier_2016}, a wind velocity of $U_a = 1$~m/s (a value in the wrinkle regime, below the transition to regular waves) gives $u^* \simeq 0.05$~m/s, and hence $Re_\delta \simeq 100$, $Bo_\delta  \simeq 15$, $Fr_\delta \simeq 0.1$ and $\tilde{\nu}_\ell \simeq 6 \times 10^{-5}$. In this regime the air flow is turbulent and excites surface deformations essentially in the gravity regime with weak viscous dissipation.  Larger values of $\delta$, as found in experiments with larger fetch and in the ocean, naturally fall in that regime too.

\subsection{Spectral formulation}

Since the surface deformations in the wrinkle regime are very small, we can neglect their feedback on the turbulent boundary layer. The problem is therefore linear and, assuming that all fields are statistically stationary and homogeneous, they can be described by their space-time Fourier transform, e.g., for the surface deformation field
\begin{equation}\label{eq:fourier_zeta}
\hat{\zeta}(\textbf{k},\omega)=\mathcal{F}\{\zeta(\textbf{r},t)\}=\int \text{d}^2 {\bf r} \text{d}t \, \zeta(\textbf{r},t)e^{-i(\textbf{k}\cdot \textbf{r}-\omega t)}
\end{equation} 
and similarly for the pressure $p(\textbf{r},t)$ and shear stress $\boldsymbol{\sigma}(\textbf{r},t)$ at the liquid surface, with $\textbf{r}=x\textbf{e}_x+y\textbf{e}_y$ and $\textbf{k}=k_x\textbf{e}_x +k_y\textbf{e}_y$ the horizontal position and wave vector, respectively.  The assumption of statistical stationarity implies that viscous dissipation balances the turbulent energy input: we therefore ignore the quasi-inviscid growth regime of Phillips~\cite{Phillips_1957} and focus on the viscous-saturated wrinkle regime.

\begin{figure}[bt]
	\includegraphics[width=1\linewidth]{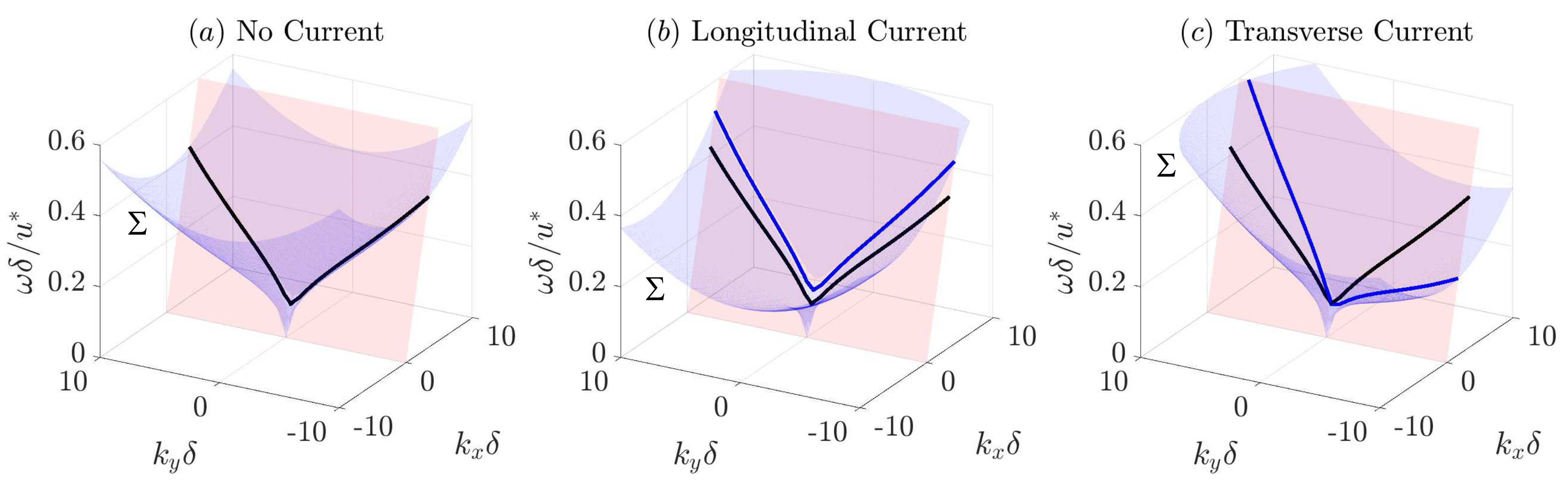}
\caption{(a) Representation in Fourier space of the resonant surface $\Sigma$, defined as ${\Re} \{ D({\bf k},\omega) \} = 0$ (blue surface),  and the forcing plane $\omega=U_c k_x$ (pink surface), where $U_c$ is the convection velocity of the source. The intersection between these two surfaces (black line) is where the energy of the surface response is expected. (b) Same representation in presence of a uniform longitudinal current $U_s / u^* = 2$. The resonant surface is now ${\Re} \{ D({\bf k},\omega - \omega_D({\bf k})) \} = 0$ (blue surface), with $\omega_D$ the Doppler shift. It intersects the forcing plane (blue line) for larger $k_x$, thus producing shorter structures. (c) Same representation in presence of a uniform transverse current $U_s / u^* = 2$. The intersection of the resonant surface $\Sigma$ with the forcing plane is now tilted (blue line), thus modifying the orientation of the wrinkles.}
	\label{fig:dispersion_bol}
\end{figure}

For laminar flow in the liquid and for small wave slopes, $\hat{\zeta}(\textbf{k},\omega)$ takes the form of a resonant response in Fourier space~\cite{Perrard2019}
\begin{equation}\label{eq:zetahat}
\hat{\zeta}(\textbf{k},\omega)=\frac{\hat{S}(\textbf{k},\omega)}{{D}(\textbf{k},\omega)},
\end{equation} 
where $\hat{S}(\textbf{k},\omega)$ is the spectral forcing related to the pressure and shear stress Fourier transform,
\begin{equation}\label{eq:source}
\hat{S}(\textbf{k},\omega)=(k\hat{p}+i\textbf{k}\cdot \boldsymbol{\hat{\sigma}})/\rho_\ell,
\end{equation} 
and ${D}(\textbf{k},\omega)$ is an inverse convolution kernel,
\begin{equation}\label{eq:dispersion}
{D} (\textbf{k},\omega) = (\omega + 2 i \nu_\ell k^2)^2-\omega_r^2(k),
\end{equation}
with $\omega_r^2(k) = (g+\gamma k^2/\rho_\ell )k$ the inviscid dispersion relation of capillary-gravity waves in infinite depth, and $k=|\textbf{k}|$. In Eq.~(\ref{eq:dispersion}) the small viscosity limit $\nu_\ell k^2 \ll \omega$ is assumed. Waves $(\textbf{k},\omega)$ satisfying ${\Re} \{ {D} \} =0$ form an axisymmetric surface noted $\Sigma$  in Fig.~\ref{fig:dispersion_bol}(a). Equation~(\ref{eq:zetahat}) shows that the energy of the surface response is significant for waves $(\textbf{k},\omega)$ excited by the forcing and matching the dispersion relation. In a turbulent boundary layer in the $x$ direction, the forcing is significant along a tilted plane of equation $\omega = k_x U_c$ (shown in pink in Fig.~\ref{fig:dispersion_bol}), with $U_c$ the characteristic convection velocity of the stress fluctuations.
This convection velocity is slightly smaller than the free-stream velocity $U_a$, with a weak dependence in wave number and Reynolds number~\cite{Willmarth_1962,Choi_1990}; here we consider $U_c \simeq 0.6 U_a$ as a representative value.  Energy of the surface response is therefore typically found along the black line, defined as the intersection between the resonant surface $\Sigma$ and the forcing plane $\omega = k_x U_c$.

\begin{figure}[tb]
	\includegraphics[width=1\linewidth]{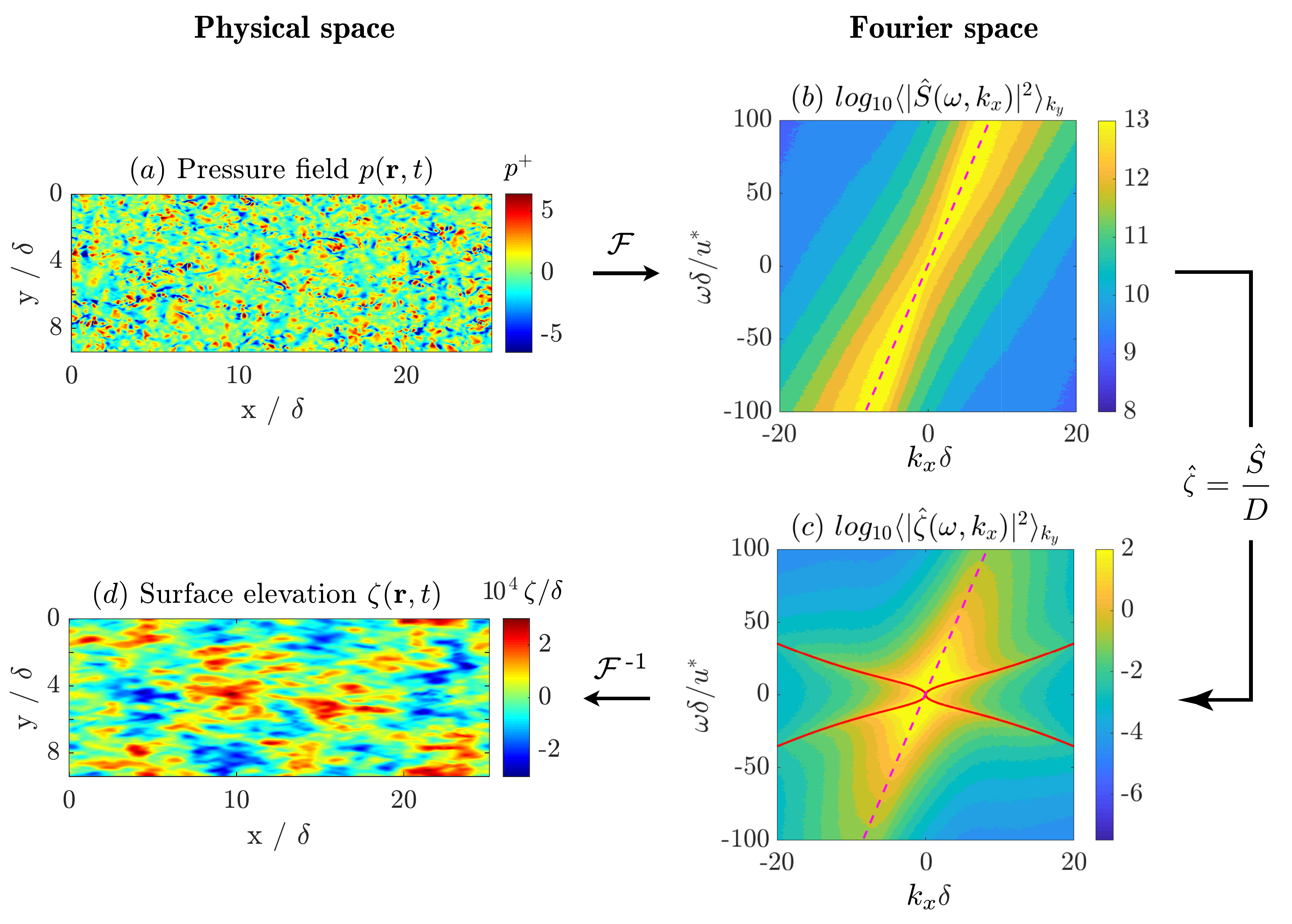}
	\caption{Illustration of the procedure used to compute the surface deformation field $\zeta({\bf r},t)$. The turbulent pressure $p({\bf r},t)$ at the liquid surface $z=0$ (a), obtained from DNS, is Fourier-transformed to compute the spectral source $\hat{S}({\bf k},\omega)$, shown in (b) in the plan ($k_x,\omega$) with average along $k_y$.  The pink dashed line shows $\omega=U_ck$, with $U_c$ the convection velocity of the pressure fluctuations. This spectral source serves as an input for the calculation of surface deformation spectrum $\hat{\zeta}({\bf k},\omega)$ using Eq.~(\ref{eq:zetahat}), illustrated in (c). The red lines show the dispersion relation $\omega_r({\bf k})$ for $k_y=0$. Finally, the surface elevation field in the physical space, shown in (d), is recovered by inverse Fourier transform (\ref{eq:zeta_real}).}
	\label{fig:method}
\end{figure} 

In Ref.~\cite{Perrard2019} we found that the shear stress contribution is negligible, and we  consider in the following only the pressure contribution, $\hat{S}(\textbf{k},\omega) = k \hat{p}(\textbf{k},\omega) / \rho_\ell$. 
The surface displacement in the physical space can then be obtained by applying the inverse Fourier transform of Eq.~(\ref{eq:zetahat}),
\begin{equation}\label{eq:zeta_real}
\zeta(\textbf{r},t)=\frac{1}{(2\pi)^3} \iiint d^2\textbf{k}d\omega \, \frac{k\hat{p}(\textbf{k},\omega) / \rho_\ell}{(\omega+2i\nu_\ell k^2)^2-\omega_r^2(k)} e^{i(\textbf{k}\cdot \textbf{r}-\omega t)} ,
\end{equation}
from which the root mean square (rms) wave amplitude $\zeta_{\mathrm{rms}} = \langle \zeta^2 \rangle^{1/2}$ is obtained using Parseval's identity, 
\begin{equation}
\label{eq:zrmsPB}
\zeta_\mathrm{rms}^2 =  \frac{1}{(2\pi)^3} \iiint d^2\textbf{k}d\omega \, \frac{k ^2 |\hat{p}(\textbf{k},\omega)|^2 / \rho^2_l}{ | (\omega+2i\nu_\ell k^2)^2-\omega_r^2(k) |^2}.
\end{equation}
	 
Equation~(\ref{eq:zeta_real}) provides a means of calculating the surface deformations under arbitrary (but statistically homogeneous and stationary) pressure forcing. The calculation steps are illustrated in Fig.~\ref{fig:method}.

Figure \ref{fig:method}(a) shows a typical snapshot of the pressure field obtained from DNS for $Re_\delta = 250$ (numerical details are provided in Sec.~\ref{sec:methods}). It shows nearly isotropic pressure patches, of typical amplitude $\rho_a u^{*2}$ and correlation length $\Lambda \simeq 250 \delta_v$, where $\delta_v = \delta / Re_\delta$ is the thickness of the viscous sublayer (we therefore have $\Lambda \simeq \delta$ for this particular value of $Re_\delta$). The correlation length is defined here from the spectral barycenter [see Eq.~(\ref{eq:barycenter}) below], which roughly corresponds to an average wavelength in the physical space.

Figure \ref{fig:method}(b)  shows  the spectral source term $\hat{S}(\textbf{k},\omega)=k \hat{p}(\textbf{k},\omega) / \rho_\ell $ in the plane $(k_x, \omega)$, averaged along $k_y$; here $k_x$ and $\omega$ are made non-dimensional using the boundary-layer length scale $\delta$ and time scale $\delta /u^*$. The energy of the source is concentrated along the line $\omega \simeq U_c k_x$ (red dashed line), with $U_c \simeq 0.6 U_a \simeq 12 u^*$ the typical convection velocity of the pressure fluctuations.

Figure \ref{fig:method}(c)  shows the spectral response, computed using Eq.~(\ref{eq:zetahat}). We can see that the energy of the surface deformations is at wave numbers smaller than for the forcing (larger scales), and is shifted towards the dispersion relation ${\Re} \{ D \}=0$ (red lines). Note that nearly all the energy actually falls near ${\Re} \{ D \}=0$, which is axisymmetric (it depends only on $k = |{\bf k}|$), but the representation in the plane $(\omega, k_x)$ with $k_y$-averaging breaks the axisymmetry and shows energy apparently far from the dispersion relation~\cite{Perrard2019}. 

Figure \ref{fig:method}(d)  finally shows a snapshot of the resulting surface deformation in the physical space, obtained from Eq.~(\ref{eq:zeta_real}).  It shows wrinkles elongated in the wind direction, of typical amplitude $\zeta_\mathrm{rms} / \delta \simeq 10^{-4}$ and correlation lengths $(\Lambda_x, \Lambda_y) \simeq (7, 3) \delta$, significantly larger than the correlation length $\Lambda \simeq \delta$ of the pressure patches from which they originate.

\subsection{Modified dispersion relation with current}
\label{sec:models}

We now include in the spectral formulation a stationary current in the liquid ${\bf U} = U(z) {\bf \hat e}_c$, uniform in the horizontal plane $(x,y)$, with possible variation of the amplitude along the depth $z$ (Fig.~\ref{fig:sketch}). Since the current may be driven by the wind itself or by any other means, we consider here a general current of arbitrary direction, making a constant angle $\theta= \cos^{-1}({\bf \hat e}_c \cdot {\bf \hat e}_x)$ with the wind.

Waves propagating in a current have their frequency modified by a Doppler shift. The simplest situation is that of a constant current $\textbf{U} = U_s \hat{\bf e}_c$ over the entire water depth, as sketched in Fig.~\ref{fig:sketch}(a) in the longitudinal case ($\theta=0$). Although not relevant for a wind-driven surface current, this simple situation may be encountered in near-shore regions, tide currents, and rivers. In addition to the five dimensionless numbers introduced in Sec.~\ref{sec:ndn}, such a uniform current introduces two additional parameters to the problem: the normalized current velocity $U_s / u^*$, and the current direction $\theta$. In this case, the Doppler shift for a wave of wave vector ${\bf k}$ simply reads
\begin{equation}\label{eq:RD_doppler}
\omega_D ({\bf k}) =\textbf{k}\cdot \textbf{U}.
\end{equation}
The surface deformation spectrum (\ref{eq:zetahat}) is therefore obtained by replacing the inviscid dispersion relation $D({\bf k},\omega)$ [Eq.~(\ref{eq:dispersion}) with $\nu_\ell=0$] by $D({\bf k},\omega - \omega_D({\bf k}))$, showing that the resonance ${\Re} \{ D \} = 0$ therefore occurs for $(\omega - \omega_D)^2 = \omega_r^2$, i.e. for $\omega = \pm \omega_r + \omega_D$.

In the following, we restrict ourselves to a current aligned with the wind ($\theta=0$), for which $\omega_D({\bf k}) = k_x U_s$, and to a transverse current ($\theta = \pi/2$), for which $\omega_D({\bf k}) = k_y U_s$. These cases are illustrated in Fig.~\ref{fig:dispersion_bol}(b) and Fig.~\ref{fig:dispersion_bol}(c), showing the Doppler-shifted dispersion relation and the resulting intersection with the forcing plane. From these figures we can anticipate that wrinkles with longitudinal current will have larger $k_x$ (finer scales), while wrinkles with transverse current will be tilted. Note that for a uniform current aligned with wind, Doppler-shifting the dispersion relation is equivalent to replacing the convection velocity $U_c$ by $U_c - U_s$, i.e. to consider the forcing in the frame of the liquid. 

The situation of a depth-varying current is more complex, because each wave vector ${\bf k}$ now perceives the current at a different depth. Motivated by experimental measurements of wind-driven currents in deep water~\cite{swan2000simple,swan2001experimental,caulliez2007turbulence,breivik2014approximate}, we consider here a simple exponential velocity profile characterized by a thickness $\delta_\ell$ and surface velocity $U_s$,
\begin{equation}\label{eq:exp_profile}
{\bf U}(z) = U_s e^{z/\delta_\ell} {\bf \hat e}_c,
\end{equation}
sketched in Fig.~\ref{fig:sketch}(b). This introduces $\delta_\ell / \delta$ as an additional dimensionless parameter in the problem. The expected effect of this sheared current is to high-pass filter the Doppler shift with a cutoff at $k\simeq \delta_\ell^{-1}$: Wavelengths smaller than $\delta_\ell$ are simply advected by the surface current, so their frequency is 
Doppler-shifted by an essentially constant velocity $U_s$, while much larger wavelengths propagate on an almost static liquid and have their frequency unchanged.

The influence of a depth-varying current on the dispersion relation has been the subject of several studies, all assuming inviscid wave propagation. A difficulty arises here in defining a relevant nondimensional measure of the shear intensity~\cite{ellingsen2017approximate}. For a given wavenumber $k$, we wish to compare the intrinsic wave frequency $\omega_r$ to the typical shear rate perceived at the scale of the wave, i.e., the shear rate $dU/dz$ at the depth $|z| \simeq k^{-1}$. For the exponential profile (\ref{eq:exp_profile}) this shear rate is essentially $U_s/\delta_\ell$ for small wavelength ($k \delta_\ell \gg 1$): the weak shear criterion is therefore $U_s / c_\ell \ll 1$, with $c_\ell = \omega_r \delta_\ell$ the phase velocity of waves of wavelength $\simeq \delta_\ell$. If the thickness $\delta_\ell$ of the current layer is comparable to the thickness $\delta$ of the turbulent boundary layer, this criterion can be expressed  in the more conventional form $U_s/c \ll 1$, where $c \simeq \omega_r \delta$ is the phase velocity of the dominant waves. This ratio $U_s/c$ is frequently used as an approximate nondimensional measure of the shear intensity, and we shall use it for simplicity in the following.

The simplest model, introduced by Stewart and Joy \cite{stewart1974hf}, modifies the dispersion relation of waves in infinite depth by a simple additive Doppler-like term, valid to first order in $U_s/c$,
\begin{equation}\label{eq:RD_SJ}
\omega_D ({\bf k}) =k\int_{-\infty}^0 2\textbf{k} \cdot \textbf{U}(z)e^{2kz}dz.
\end{equation}
A finite-depth extension was later proposed by Skop \cite{skop1987approximate} that was then developed to second order by Kirby and Chen \cite{kirby1989surface}. The case of a sheared current with both amplitude and direction varying with $z$ in finite depth was recently analyzed for small curvature of $U(z)$~\cite{ellingsen2017approximate}, and generalized to arbitrary current and depth variations by Li and Ellingsen~\cite{LiEllingsen2019}. Here we restrict to weak currents of varying amplitude but constant direction in infinite depth. Interestingly, the first-order development (\ref{eq:RD_SJ}) of Stewart and Joy \cite{stewart1974hf} is almost indistinguishable from the exact solution even for $U_s/c \simeq O(1)$~\cite{ellingsen2017approximate}. Since we have $u^*/c\simeq O(1)$, this condition is satisfied in the following for currents $U_s / u^* \simeq O(1)$.

Until now, the effects of viscosity have been ignored. Investigating the influence of a sheared current on the wrinkle properties is challenging, due to the combined effect of viscosity and shear which must be taken into account in the spectral formulation. Although the general case with finite viscosity and finite current has not been considered in the literature, we can infer the form of the modified dispersion relation if we assume that both quantities are small. To linear order in both $U_s/c$ and $\tilde{\nu}_\ell$, the individual corrections simply add up, giving the modified dispersion relation
\begin{equation}\label{eq:D_modified}
D({\bf k},\omega) = [\omega-\omega_D({\bf k}) + 2 i \nu_\ell k^2]^2 - \omega_r^2(k),
\end{equation} 
with $\omega_D({\bf k})$ given by Eq.~(\ref{eq:RD_SJ}). We note that this form satisfies the Hermitian symmetry of the problem: The surface deformation being real, its Fourier transform satisfies $\hat \zeta (-{\bf k}, -\omega) = \hat \zeta^\dagger({\bf k}, \omega)$, with $^\dagger$ the complex conjugate, and so does $D({\bf k},\omega)$.

In the following, we investigate the influence of the three dimensionless numbers, $U_s/u^*$, $\delta_\ell/\delta$ and $\theta$, on the wrinkle properties. We focus only on the extreme cases of purely longitudinal ($\theta=0$) and transverse ($\theta=\pi/2$) currents. We naturally expect that, for a given current amplitude $U_s$,  the most pronounced effects on wrinkles are for a uniform profile, i.e., for $\delta_\ell / \delta \gg 1$, which equally affects all wave vectors.  On the other hand, since the characteristic wavelengths of the wrinkles are of order $\delta$, we expect vanishing effects in the limit  $\delta_\ell / \delta \ll 1$ (thin flowing layer on a liquid at rest). For this reason, we first consider the upper limit $\delta_\ell / \delta \gg 1$, before studying the more realistic case of finite $\delta_\ell / \delta$.

\section{Numerical methods}
\label{sec:methods}

\subsection{DNS simulations}

We now describe the dynamics of the surface deformations forced by the turbulent boundary layer in the air. We follow here the simplified one-way approach introduced in Perrard {\it et al.}~\cite{Perrard2019}: we neglect the feedback of the waves on the dynamics of the turbulent boundary layer in the air. We can therefore use a data base of time-resolved pressure fields  extracted from DNS of a turbulent channel flow  with flat walls and no-slip boundary conditions. The channel half-height corresponds to the boundary layer thickness $\delta$, and periodic boundary conditions are applied in the streamwise and spanwise directions.  Table \ref{tab:DNS_data} summarizes the DNS parameters used for the different cases, with $Re_{\delta}$ ranging from 100 to 550.

Assuming a no-slip boundary condition at the interface instead of the true velocity and stress continuity is discussed in Ref.~\cite{Perrard2019} in the absence of current. It was shown that this simplification is acceptable in the wrinkle regime, i.e., for small wave amplitude and wave slope. Extending this assumption in the presence of a sheared current is justified because the convection velocity of the stress fluctuations, $U_c \simeq 12 u^*$, is much larger than the surface velocity $U_s \simeq u^*$ considered here. The flow in the liquid being assumed laminar, the current in the liquid (driven by the mean component of the shear stress) and the wrinkles (excited by the fluctuating component of the stresses) can be considered separately. Only the dynamics of the wrinkles is computed, while the sheared current is considered as prescribed, and acts only through the modification of the dispersion relation.

We compute the source term ${\hat S} (\textbf{k},\omega)$ from the space-time Fourier transform of the wall pressure on a discrete three-dimensional Cartesian grid $(k_x,k_y,\omega)$.
The size of the computational box $L_x \times L_y$  must be carefully chosen to ensure a sufficient spectral resolution to allow evaluation of the surface deformation spectrum. The minimum channel size $(2\pi,\pi)\delta$ often used in turbulent channel flows is not sufficient here for the study of wrinkles: while pressure fluctuations within the turbulent boundary layer are dominated by the (inner) viscous sublayer thickness $\delta_{\nu}$, this is not the case for wrinkles, which are dominated by the (outer) boundary layer thickness $\delta$. This is because the surface response shifts the supplied energy to smaller $k$ (larger scales), yielding a maximum energy at the upper bound $\delta$ of the forcing interval~\cite{Perrard2019}: wrinkles are therefore highly sensitive to the small energy content of the pressure fluctuations at the largest scales, which must be correctly resolved.  Here we use boxes of size $(8\pi, 3\pi)\delta$ and $(60\pi, 6\pi)\delta$.  The largest box resolves almost all the energy spectrum: structures up to half the box length contain more than 80\% of the energy~\cite{lozano2014effect}.  However, due to the high computational cost, only the lowest Reynolds number ($Re_\delta = 100$) is available for this largest box, whereas higher $Re_\delta$ are available for the intermediate $(8\pi, 3\pi)\delta$ box only.

In the following, the other dimensionless numbers are chosen as follows: $\rho_a / \rho_\ell = 1.2 \times 10^{-3}$ (air-water density ratio), $Bo_\delta = 14$ (waves forced essentially in the gravity regime), and a normalized liquid viscosity in the range $\tilde{\nu}_\ell = \nu_\ell / \sqrt{g \delta^3} \simeq 6 \times 10^{-5} - 6 \times 10^{-3}$. For a boundary-layer thickness $\delta = 3$~cm such as in the experiments of Paquier {\it et al.}~\cite{Paquier_2015,Paquier_2016}, this range covers 1--100 times the viscosity of water.

\begin{table}[!t]
\caption{\label{tab:DNS_data} Details of the DNS turbulent channel air flow for the different Reynolds numbers $Re_{\delta}=u^*\delta/\nu_a$. $\Delta x^+$ and $\Delta y^+$ are the spatial resolutions in terms of Fourier modes before dealisasing (in wall units, normalized by $\delta_\nu = \nu_a/u^*$). $\Delta z_\mathrm{min}^+$ and $\Delta z_\mathrm{max}^+$ are the finest and coarsest spatial resolutions in the wall-normal direction. $\Delta t^+$ is the temporal separation between stored flow fields (in units of $\delta_\nu/u^*$) and $T_\mathrm{max}$ is the total duration of the simulation.}
\begin{tabular}{p{2.3cm}p{1.5cm}p{1.5cm}p{1.5cm}p{1.5cm}p{1.5cm}p{1.5cm}p{1.5cm}}
\hline \hline
   Box size & Re$_{\delta}$ & $\Delta x^+$ & $\Delta y^+$ & $\Delta z_\mathrm{min}^+$ & $\Delta z_\mathrm{max}^+$ & $\Delta t^+$ & $T_\mathrm{max} u^*/\delta$ \\ \hline
   $(8\pi, 3\pi)\delta$ & 100 & 10.1 & 5.7 & 0.06 & 3.4 & 0.63 & 12.5 \\
   & 180 & 9.1 & 5.3 & 0.02 & 3.0 & 0.64 & 14.1 \\
   & 250 & 12.1 & 6.8 & 0.03 & 4.0 & 0.61 & 10.1\\
   & 360 & 13.1 & 6.5 & 0.04 & 5.8 & 3.80 & 21.8\\
   & 550 & 13.4 & 7.5 & 0.04 & 6.7 & 0.45 & 6.7\\
   $(60\pi, 6\pi)\delta$ & 100 & 9.5 & 7.3 & 0.06 & 3.4 & 0.63 & 50.5\\
\hline \hline
\end{tabular}

\end{table}

\subsection{Spectral interpolation method}
\label{sec:interpolation}

\begin{figure}[tb]
	\includegraphics[width=0.9\linewidth]{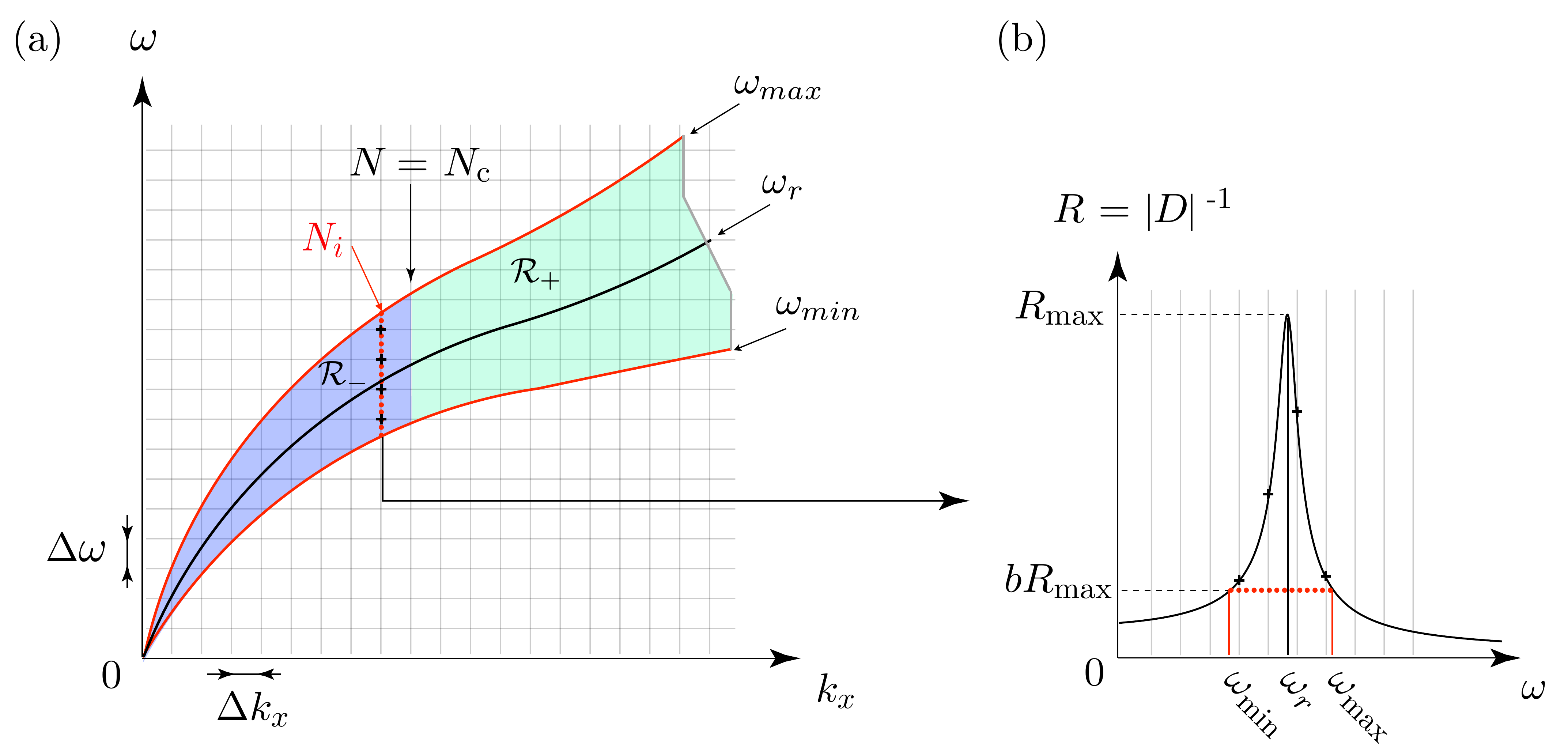}
	\caption{Illustration of the spectral interpolation method. (a) The inviscid dispersion relation $\omega = \omega_r({\bf k})$ is plotted in the plane $k_y=0$, surrounded by the resonant subspace bounded by $\omega_\mathrm{min}$ and $\omega_\mathrm{max}$ (shaded areas). The resonant subspace is defined such that $R({\bf k}, \omega) > b R_\mathrm{max}({\bf k})$, with $R_\mathrm{max}({\bf k})$ the maximum of the resonance function, as shown in (b). The spectral grid is represented in gray, with mesh sizes $\Delta \omega = 2\pi / T_\mathrm{max}$ and $\Delta k_x = 2\pi / L_x$.  The resonant subspace is split between an under-resolved subspace ${\cal R}_-$  such that $N<N_c$ (in blue), and a resolved subspace ${\cal R}_+$ such that $N>N_c$ (in green), with $N$ the number of points along $\omega$ and $N_c$ a threshold. The grid is refined along $\omega$ in the under-resolved subspace ${\cal R}_-$ up to a total number of $N_i$ points (in red). The evaluation of $\hat{\zeta}$ in this interval is performed by applying a linear interpolation of the source term $\hat{S}(\textbf{k},\omega)$ on the refined grid.}
	\label{fig:interpolation}
\end{figure}

A strong numerical constraint when computing the space-time Fourier transform $\hat \zeta(\textbf{k},\omega)$ from Eq.~(\ref{eq:zetahat}) arises from the small thickness of the resonance around the dispersion relation, which may be below the spectral resolution  $\Delta k_{(x,y)}  = 2\pi/L_{(x,y)}$ and $\Delta \omega = 2\pi/T_\mathrm{max}$ if the box size $(L_x, L_y)$ and time duration $T_\mathrm{max}$ of the sample are too small. To evaluate the thickness of the dispersion relation, we introduce the resonance function
\begin{equation}\label{eq:resonance}
R({\bf k}, \omega) = \frac{1}{\vert {D} ({\bf k}, \omega) \vert}=\frac{1}{\sqrt{(\omega^2-\omega_r^2)^2+\omega_{\nu}^2\omega^2}},
\end{equation}
with $\omega_{\nu}=4\nu_\ell k^2$. The effect of the current is not included here for simplicity, but it can be simply included by replacing $\omega$ by $\omega-\omega_D({\bf k})$. For a given wave vector ${\bf k}$, the maximum $R_\mathrm{max}(k)=1/(\omega_{\nu}\omega_r)$ is at $\omega=\omega_r(k)$, on the resonant surface $\Sigma$, and the typical thickness is $\omega_\nu$ [see Fig.~\ref{fig:interpolation}(b)].
The rapid variations of $R$ near its maximum, typically in the interval $[\omega_r - \omega_\nu, \omega_r + \omega_\nu]$, make the integrated product $R(\textbf{k},\omega)\hat{S}(\textbf{k},\omega)$ highly sensitive to the mesh size $\Delta \omega$, or to the exact positions of $\Sigma$ on the spectral grid.  Although a direct integration method is sufficient at large viscosity, this represents a severe limitation at small viscosity.  The smallest resolved viscosity can be estimated by equating the spectral mesh size $\Delta \omega$ and the resonance thickness $\omega_\nu$. Considering that the dominant energy is at $k \simeq \delta^{-1}$,  the smallest resolved liquid viscosity is $\nu_{\ell, \mathrm{min}} \simeq \delta^2/T_\mathrm{max}$. In terms of normalized liquid viscosity, the criterion $\tilde{\nu}_{\ell, \mathrm{min}} = \sqrt{\delta/g} / T_\mathrm{max} \ll 1$ requires a sample duration much larger than the period of the slowest gravity waves of wavelength of the order of $\delta$.

Since at small viscosity the thickness of the resonance is smaller than the thickness of the spectral forcing, 
we can overcome the limited spectral resolution by evaluating the resonance on a finer grid on which we interpolate the forcing. Here the thickness of the forcing in the Fourier space, visible in Fig.~\ref{fig:method}(b), is related to the temporal coherence of the pressure fluctuations traveling in the boundary layer. To limit the computational cost, this mesh refinement is performed only in the vicinity of the resonance, as sketched in Fig.~\ref{fig:interpolation}. For each wave vector ${\bf k}$, we define the resonant interval $[\omega_\mathrm{min},\omega_\mathrm{max}]$ surrounding the resonance $\omega_r(k)$ such that $R({\bf k}, \omega) > b R_\mathrm{max}({\bf k})$, with $b < 1$ (red boundaries in Fig.~\ref{fig:interpolation}), and count the number $N$ of mesh points in the interval (black crosses). If $N$ is smaller than a threshold value $N_c$, we refine the grid by introducing $N_i$ points in the interval $[\omega_\mathrm{min},\omega_\mathrm{max}]$ (red points).  The under-resolved resonant subspace ${\cal R}_-$ where this refinement is performed is colored in blue in Fig.~\ref{fig:interpolation},  while the resolved subspace  ${\cal R}_+$ is in green. Finally, we linearly interpolate the source $\hat{S}(\textbf{k},\omega)$ on the refined grid in ${\cal R}_-$ and compute the space-time Fourier transform $\hat{\zeta}({\bf k},\omega)$. From this refined piecewise spectrum the main spectral quantities characterizing the wrinkles can be computed with a better accuracy than from the original spectrum. The main drawback of this method is that computing the surface deformation $\zeta({\bf r},t)$ in the physical space by inverse Fourier transform is no longer possible by usual FFT algorithms, since this piecewise spectrum is not defined on a complete regular Cartesian grid.

Convergence tests were performed in order to ensure the validity of the method and determine the optimal values for the various parameters (threshold $b$, minimum number of points $N_c$ for interpolation, and number of interpolated points $N_i$). These tests were performed for different liquid viscosities and for the small and large DNS box sizes. Given that convergence was always reached for $N_i \geq 100$, we take $N_i = N_c = 100$ in the following (choosing $N_i = N_c$ ensures that there are at least $N_c$ points for each ${\bf k}$ in the resonant subspace). We choose a threshold $b=0.1$, therefore covering 90$\%$ of the resonant subspace for each ${\bf k}$. A smaller threshold would widen the selected resonant subspace, thereby implying an increase in $N_i$ and therefore in the computational cost.

\section{Influence of the current on the wrinkle properties}
\label{sec:doppler}

\subsection{Qualitative description}

\begin{figure}[!t]
	\includegraphics[width=1\linewidth]{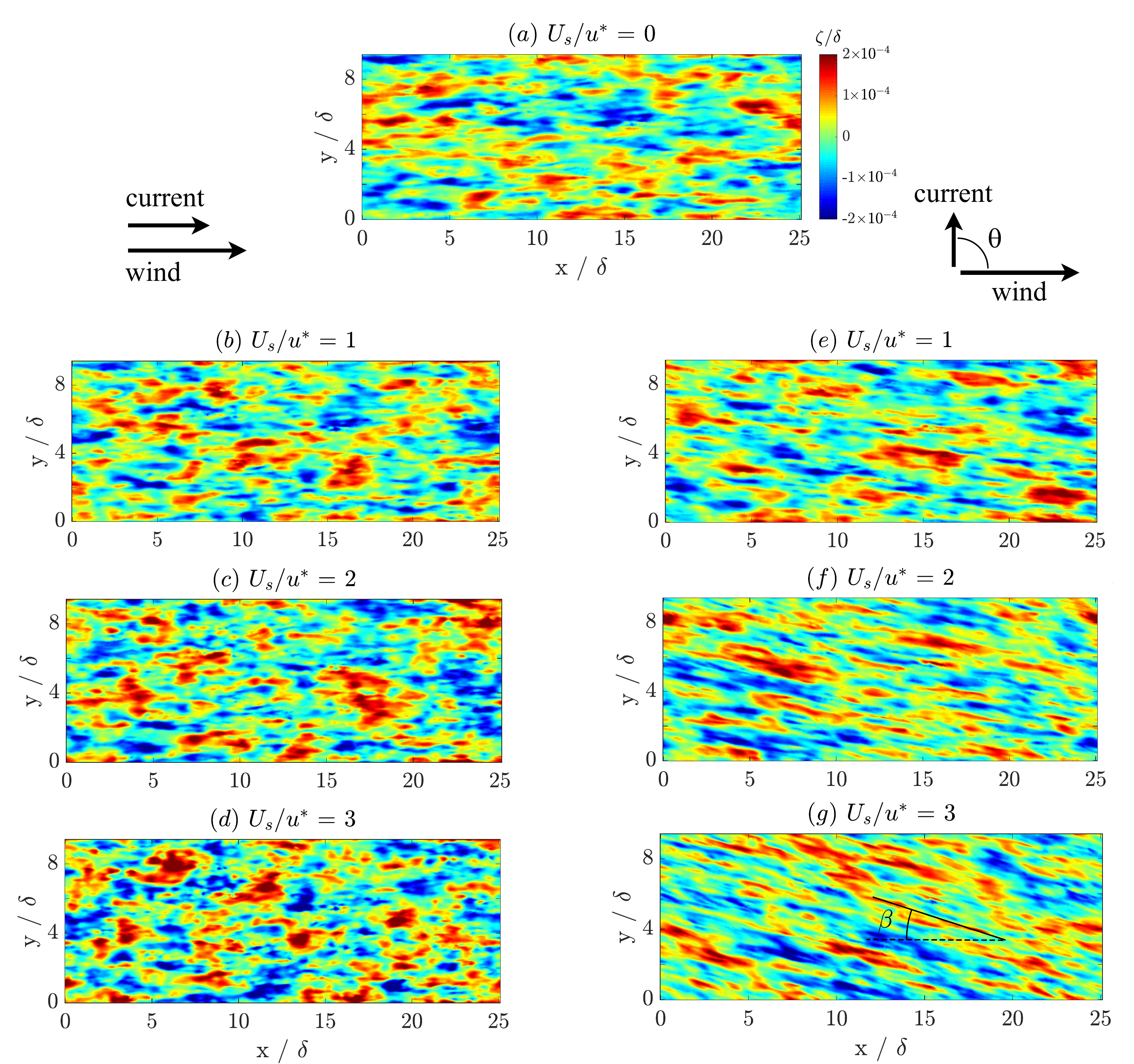}
	\caption{Surface deformations without (a) and with (b-g) current. Snapshots are compared for increasing current $U_s$ in the range $(1-3)u^*$, for a uniform current in the longitudinal (b-d) and transverse (e-g) directions ($\theta =0$ and $\theta = \pi/2$, respectively). Results are shown for $Re_{\delta}$=350 and a liquid viscosity $\tilde{\nu_\ell}$ = 6~10$^{-3}$.}
	\label{fig:visu_currents}
\end{figure}

We now analyze the overall effect of a current on the geometry of the wrinkles. Snapshots of the surface deformation $\zeta({\bf r},t)$  are shown in Fig.~\ref{fig:visu_currents} for $Re_{\delta}$ = 350, for both a longitudinal current [Figs.~\ref{fig:visu_currents}(b), \ref{fig:visu_currents}(c) and \ref{fig:visu_currents}(d), on the left-hand side] and a transverse current [Figs.~\ref{fig:visu_currents}(e), \ref{fig:visu_currents}(f) and \ref{fig:visu_currents}(g), on the right-hand side], and are compared to the reference case without current [Fig.~\ref{fig:visu_currents}(a)]. To produce these snapshots in the physical space we had to use the direct Fourier computation (\ref{eq:zeta_real}) without the spectral interpolation method of Sec.~\ref{sec:interpolation}. For this reason, we restrict our analysis here to a relatively large liquid viscosity ($\tilde{\nu_\ell}$ = 6~10$^{-3}$) to avoid discretization errors.

\begin{figure}[tb]
	\includegraphics[width=0.45\linewidth]{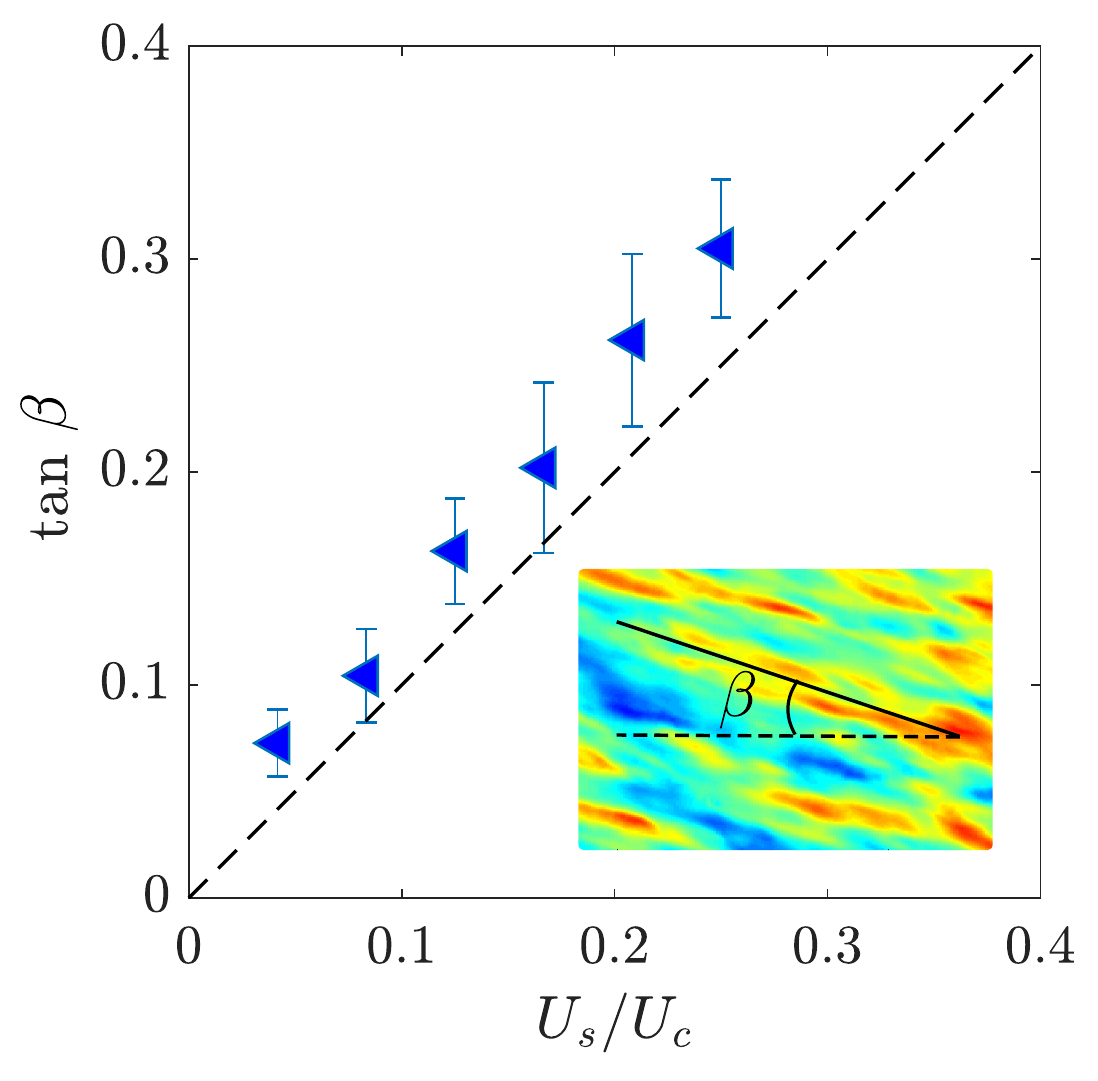}
	\caption{Tilt angle $\beta$ of the wrinkles in a transverse current at fixed Reynolds numbers $Re_{\delta}$ = 550 as a function of $U_s/U_c$. Each point is obtained by fitting straight lines through the surface deformation patterns and averaging over a large number of realizations. The dashed line is the geometric prediction $\tan \beta = U_s/U_c$, with $U_c \simeq 12 u^*$ the convection velocity of the pressure fluctuations.}
	\label{fig:angles}
\end{figure}

In the case of a transverse current, the overall shape of the wrinkles is similar to the reference case, except that they are inclined  with an angle $\beta$ that increases with the current. This angle simply reflects the sweeping by the transverse current at velocity $U_s$ of the wake behind the pressure fluctuations traveling at velocity $U_c$, yielding $\tan \beta \simeq U_s/U_c$. This relationship is in good agreement with the measured tilt angle $\beta$ shown in Fig.~\ref{fig:angles}, obtained by fitting lines through the surface deformation pattern. Note that this simple geometric construction holds only at sufficiently large Froude number, when the aperture angle of the V-shaped wakes with respect to the disturbance trajectory is itself small compared to $\beta$, i.e., when the wrinkles are sufficiently elongated~\cite{Rabaud_2013,Noblesse2013,Ellingsen2014,Darmon_2014}. This condition is satisfied in the case $Re_\delta=550$ shown here: the Froude number based on the pressure size $\Lambda$ and convection velocity $U_c$ is $Fr=U_c / \sqrt{g \Lambda} \simeq O(10)$, for which the wake essentially reduces to a line behind the disturbance. At smaller $Re_\delta$ (hence smaller $Fr$), the wake aperture is close to the Kelvin's angle of 39$^\mathrm{o}$, leading to an intricate pattern from which we cannot define a clear tilt angle $\beta$.

\begin{figure}[tb]
	\includegraphics[width=1\linewidth]{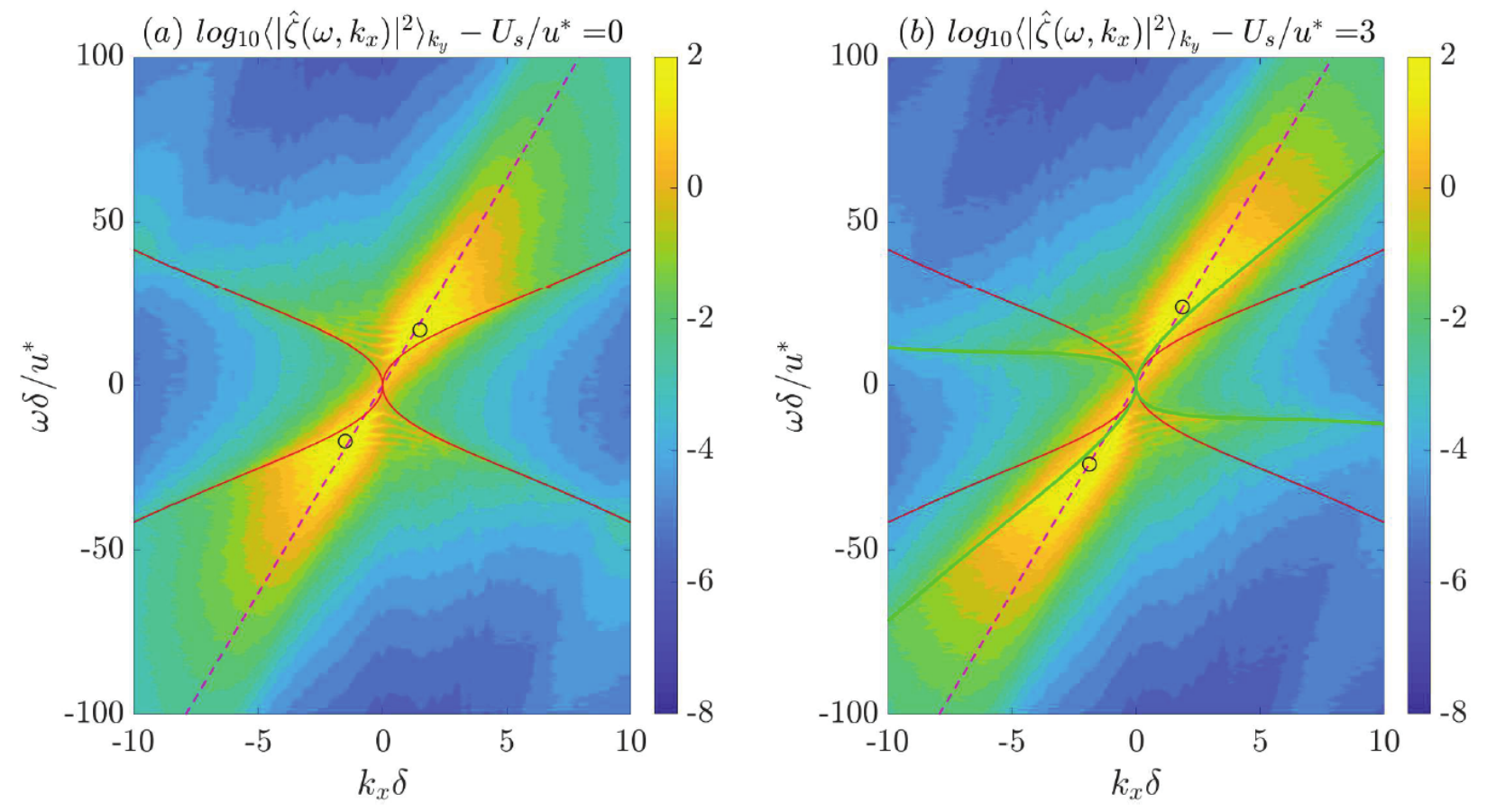}
	\caption{Space-time spectrum of the surface displacement $\vert \hat{\zeta}(k_x,\omega)\vert^2$ averaged in $k_y$ computed from Eq.~(\ref{eq:zetahat}) for liquid viscosity of $\tilde{\nu_\ell}$ = 2~10$^{-3}$ and $Re_{\delta}$ = 100, without current (a) and with a current $U_s/u^*=3$ (b). The pink dashed line shows the forcing $\omega=U_ck$, where $U_c$ is the convection velocity of the pressure fluctuations. The continuous lines represent the dispersion relation without current ($\pm \omega_r$, in red) and with a uniform current ($\pm \omega_r + \omega_D$, in green). The circles show the spectral barycenter $(K_x, \Omega)$.}
	\label{fig:spectrum_doppler}
\end{figure}

The case of a longitudinal current is more subtle. The wrinkles now remain aligned with the wind, but 
they become shorter and more fragmented as the current velocity $U_s$ is increased. This effect was expected from  Fig.~\ref{fig:dispersion_bol}(b): the Doppler-shifted dispersion relation becomes closer to the spectral forcing plane as $U_s$ is increased, therefore exciting a larger range of wavenumbers.   This is confirmed by the space-time spectrum of the surface response $\vert \hat{\zeta} \vert^2$ averaged along $k_y$ in Fig.~\ref{fig:spectrum_doppler}, which shows a clear accumulation of energy along the Doppler-shifted dispersion relation ($\pm \omega_r + \omega_D$, in green) as it becomes closer to the spectral forcing ($k_x U_c$, in dotted lines); we recall here that the energy away from the dispersion relation is an artefact of the averaging over $k_y$, which respects the symmetry of the
source but not that of the dispersion relation (see Fig.~\ref{fig:dispersion_bol}).

\begin{figure}[tb]
	\includegraphics[width=0.5\linewidth]{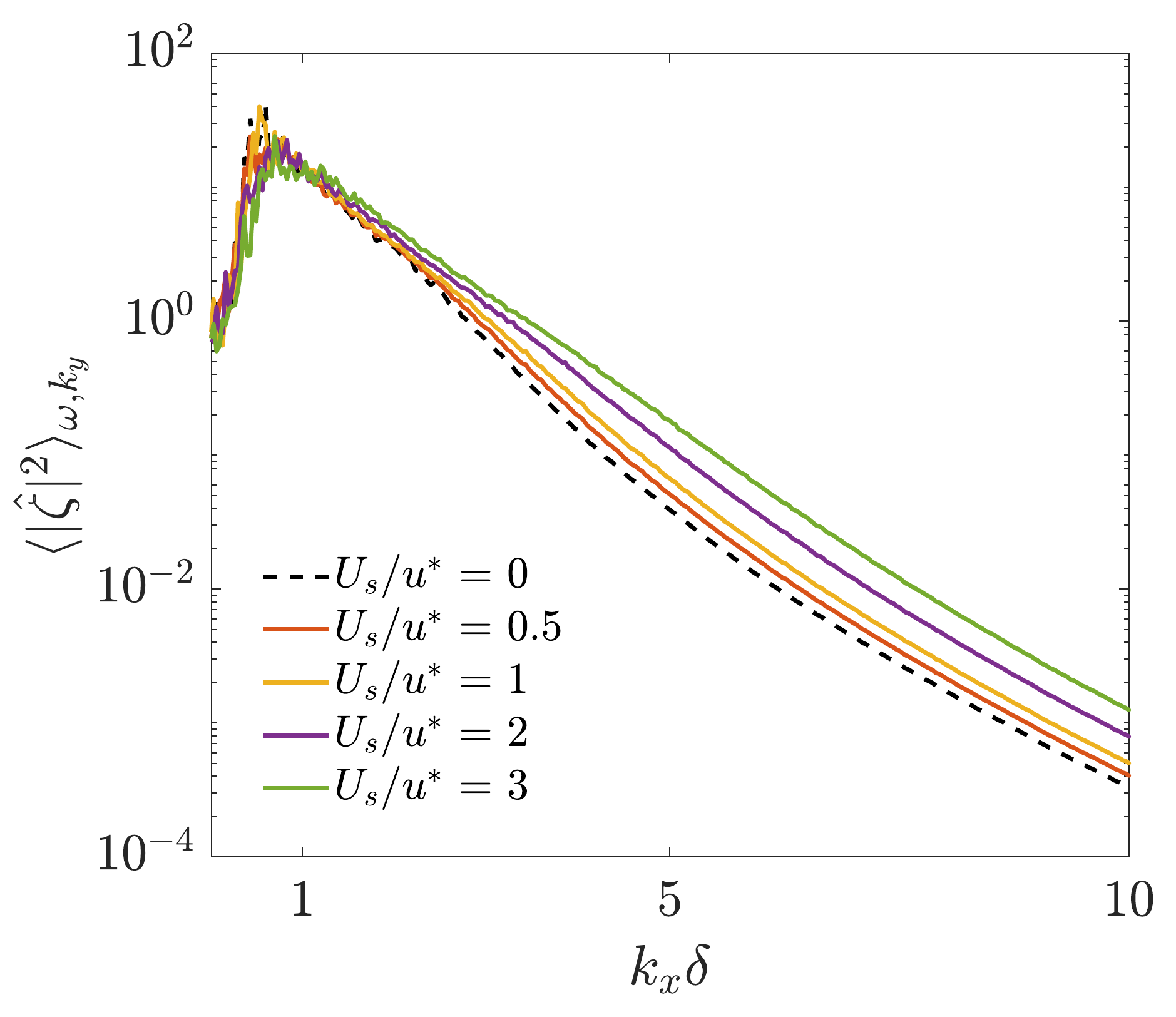}
	\caption{One-dimensional energy spectrum $|\hat \zeta|^2$ of the surface deformation averaged over $\omega$ and $k_y$ for increasing current $U_s$ in the longitudinal direction. Results are obtained for $Re_{\delta}=100$ in the large DNS box $(60\pi, 6\pi)$ with liquid viscosity $\tilde{\nu_\ell}$ = 6~10$^{-3}$.}
	\label{fig:1D_spectrum_currents}
\end{figure}

The wider range of excited wave numbers in the presence of a longitudinal current is evident in the one-dimensional spectrum $E(k_x) = \langle | \hat \zeta|^2 \rangle_{\omega, k_y}$ shown in Fig.~\ref{fig:1D_spectrum_currents}, obtained by averaging the space-time spectrum $\langle | \hat \zeta|^2 \rangle_{k_y}$ of Fig.~\ref{fig:spectrum_doppler} over $\omega$. As the current velocity $U_s$ is increased, the spectra show wider tails, with up to five times more energy at large $k_x$ for the strongest current $U_s/u^* = 3$. However,  the peak of the spectrum remains around $k_x \delta \simeq 1$, corresponding to wrinkle length $\Lambda_x = 2\pi / k_x \simeq 6 \delta$, suggesting a weak influence of the current on the energy-containing scale of the wrinkles. This weak influence is better characterized by the spectral barycenters of the wave vector and frequency,
\begin{equation}\label{eq:barycenter}
\textbf{K}=K_x {\bf \hat e}_x + K_y {\bf \hat e}_y=\frac{\int_\mathcal{D} d^2\textbf{k}d\omega \, \textbf{k}\vert \hat{\zeta}\vert ^2}{\int_\mathcal{D} d^2\textbf{k}d\omega\vert \hat{\zeta}\vert ^2}
\end{equation}
and
\begin{equation}\label{eq:freq_barycenter}
\Omega=\frac{\int_\mathcal{D} d^2\textbf{k}d\omega \, \omega\vert \hat{\zeta}\vert ^2}{\int_\mathcal{D} d^2\textbf{k}d\omega\vert \hat{\zeta}\vert ^2},
\end{equation}
where $\mathcal{D}$ is the domain of integration, $k_{x,y}>0$. The spectral barycenter $(K_x, \Omega)$, represented by black circles in Fig.~\ref{fig:spectrum_doppler}, is indeed shifted towards larger $k_x$ with current, but this shift remains moderate.

\begin{figure}[tb]
\includegraphics[width=1\linewidth]{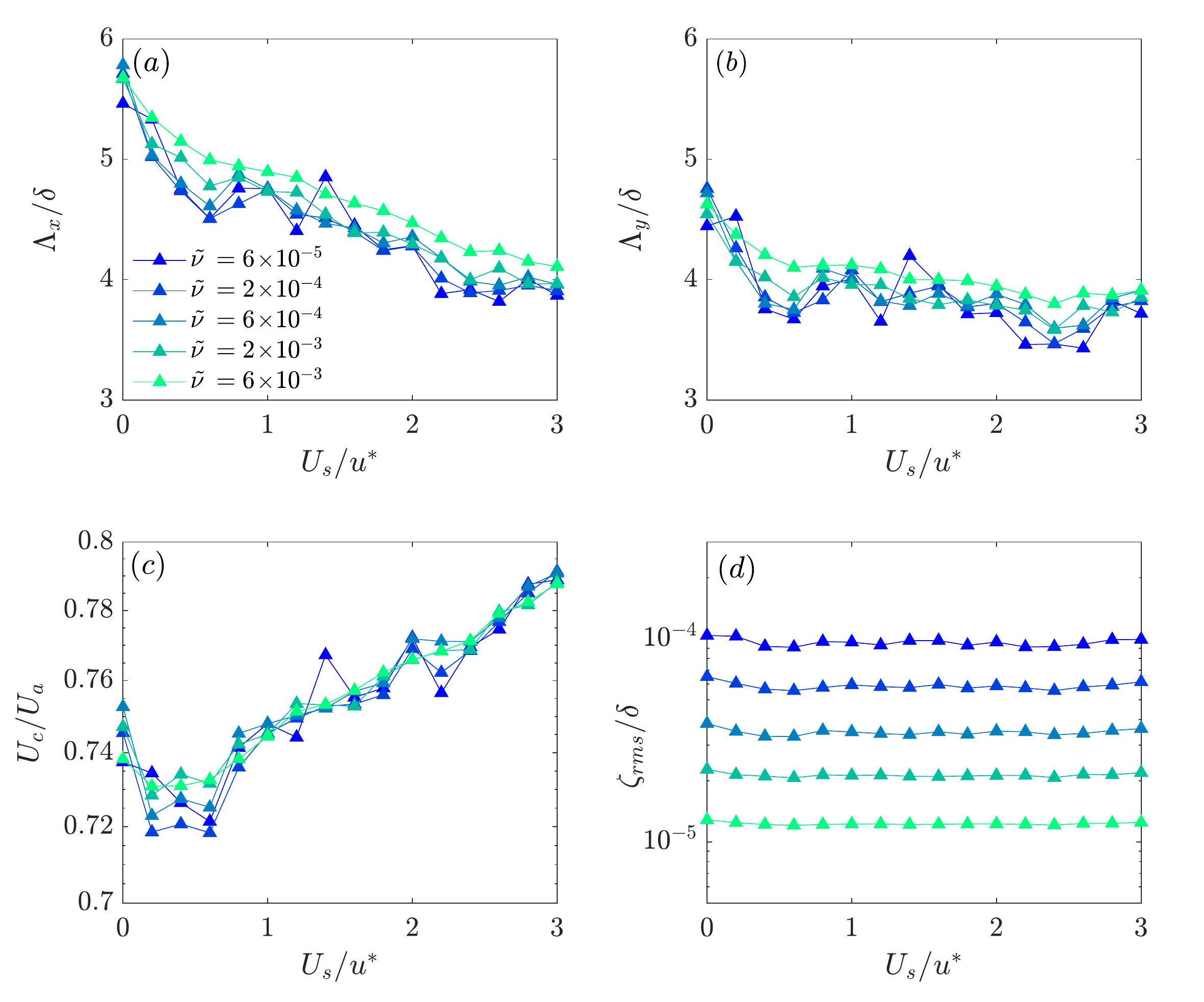}
	\caption{Modification of the wrinkle properties for a longitudinal current as a function of $U_s/u^*$: 	Characteristic streamwise $\Lambda_x/\delta$ (a) and spanwise $\Lambda_y/\delta$ (b) lengths, convection velocity $U_c/U_a$ (c), and wrinkle amplitude $\zeta_\mathrm{rms}/\delta$ (d). Results are obtained for $Re_{\delta}=100$ in the large DNS box $(60\pi, 6\pi)$, with liquid viscosity  varied in the range $\tilde{\nu_\ell} = 6~10^{-5} - 6~10^{-3}$.}
	\label{fig:wrinkles_doppler}
\end{figure}

\subsection{Wrinkle properties in a longitudinal current}
\label{sec:wplc}

In the following we systematically characterize the influence of the current on the wrinkle properties using the following four quantities: the longitudinal and transverse scales, defined from the spectral barycenter (\ref{eq:barycenter}) as $\Lambda_x=2\pi/K_x$ and $\Lambda_y=2\pi/K_y$, the wrinkle characteristic velocity $U_c = \Omega / K_x$, and the wrinkle rms amplitude (\ref{eq:zrmsPB}).
To decrease the viscosity down to conditions relevant to air-water applications ($\tilde{\nu_\ell} = 6~10^{-5}$ for $\delta \simeq 3$~cm), we now apply the spectral interpolation method described in Sec.~\ref{sec:interpolation}, and first restrict our analysis to the smallest Reynolds number $Re_\delta = 100$, for which the large DNS box $(60\pi, 6\pi)$ is available.

Figure \ref{fig:wrinkles_doppler} presents the four wrinkle properties $\Lambda_x/\delta$, $\Lambda_y/\delta$, $U_c/U_a$ and $\zeta_\mathrm{rms}/\delta$ as a function of the normalized current $U_s/u^*$, for various liquid viscosities in the range $\tilde{\nu_\ell}$ = 6~10$^{-5}$ to 6~10$^{-3}$. We first note that the length scales $\Lambda_x$ and $\Lambda_y$ show no significant dependence in $\tilde{\nu}_\ell$, whereas the wrinkle amplitude $\zeta_\mathrm{rms}$ decreases as $\tilde{\nu}_\ell^{-1/2}$, in agreement with Eq.~(\ref{eq:wab}). These scalings confirm the analytical predictions of Perrard {\it et al.}~\cite{Perrard2019} derived in the limit of small viscosity. In spite of our spectral interpolation method, results still show some noise at small $\tilde{\nu}_\ell$: the curves obtained for the lowest viscosity, for which the resonance is below the spectral resolution, show residual fluctuations of about 5$\%$ (without the spectral interpolation method the fluctuations are typically 10 times larger so that only results at large viscosity would be reliable).

The main result of Fig.~\ref{fig:wrinkles_doppler} is that the amplitude of the wrinkles is independent of the current $U_s$, whereas their characteristic sizes and convection velocity are slightly modified. Best linear fits yield
\begin{subequations}
\begin{align}
\label{eq:fitLx}
\Lambda_x/\Lambda_{x0} &\simeq 1-(0.08\pm 0.02) \, U_s / u^*,\\ 
\Lambda_y/\Lambda_{y0} &\simeq 1-(0.04\pm 0.02) \, U_s / u^*,\\ 
U_c/U_{c0} &\simeq 1+(0.03\pm 0.01) \, U_s / u^*,
\end{align}
\end{subequations}
where the subscript '$_0$' denotes the reference values without current.  These dependencies are clearly limited, confirming that the wrinkle properties are robust with respect to currents. The strongest dependence is for the streamwise size $\Lambda_x$, which decreases by 8\% for a current $U_s / u^* = 1$. This decrease of $\Lambda_x$ can be qualitatively recovered from the match between the forcing $k_x U_c$ and the Doppler-shifted inviscid dispersion relation for gravity waves, $\sqrt{g k} + k_x U_s$, yielding for $k=k_x$
\begin{equation}
\Lambda_x/\Lambda_{x0}=1-\frac{u^*}{U_c}\frac{U_s}{u^*},
\end{equation}
with $ u^* / U_c \simeq 0.08$ at $Re_{\delta}$ = 100, in good agreement with Eq.~(\ref{eq:fitLx}). The convection velocity of the wrinkle increases with surface current, but here again by a very limited amount, 3\% for $U_s / u^* = 1$.

\subsection{Influence of the Reynolds number and current thickness}
\label{sec:wpsc}

\begin{figure}[tb]
	\includegraphics[width=0.9\linewidth]{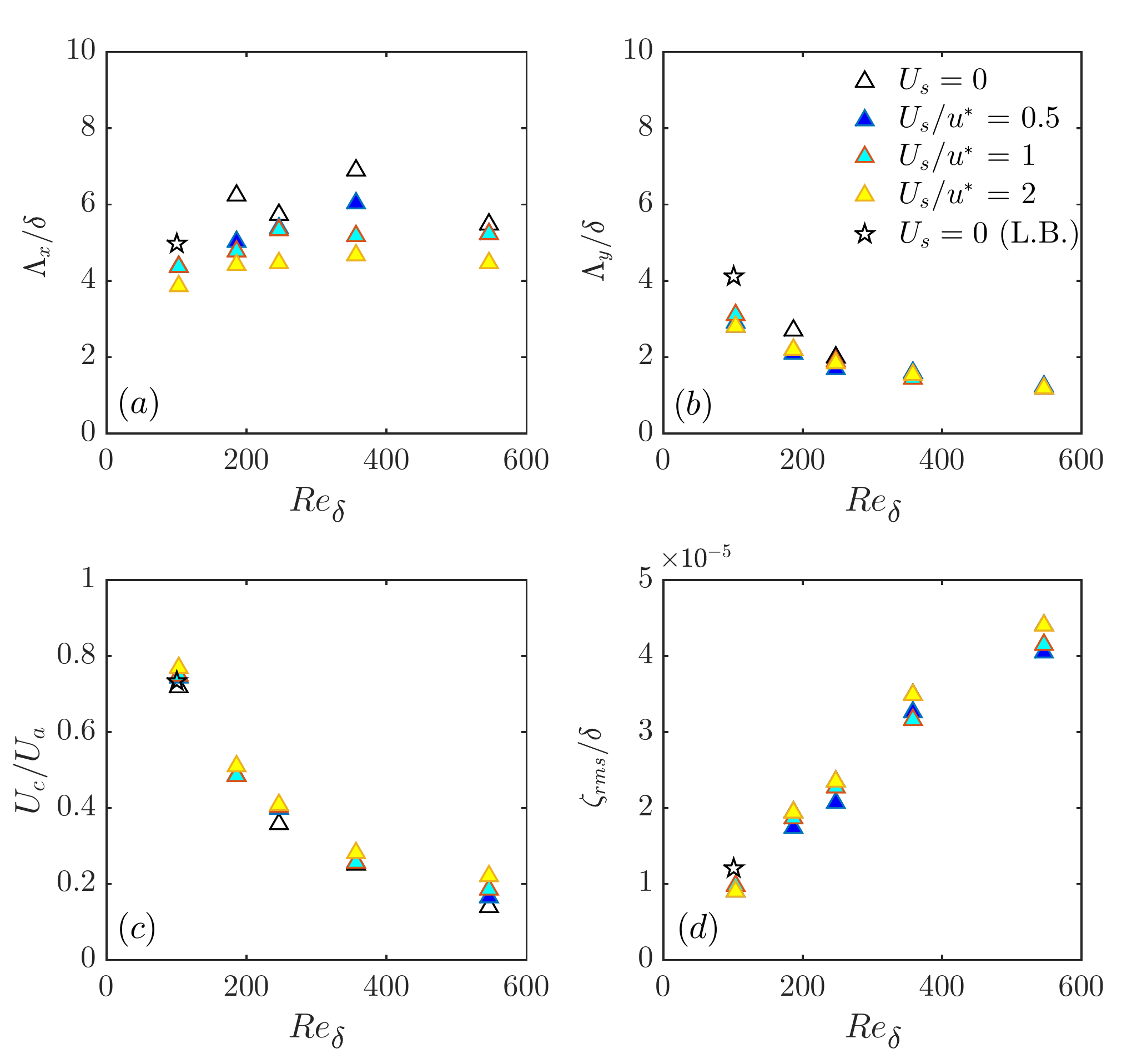}
\caption{Wrinkle properties as a function of the Reynolds number $Re_{\delta}$ for three different currents $U_s/u^*$ and a liquid viscosity  $\tilde{\nu_\ell}$ = 6~10$^{-3}$. Results are obtained using the spectral interpolation method described in Sec.~\ref{sec:methods}. $\triangle$: small box, for various values of $U_s/u^*$; $\star$: large box (L.B.), for $U_s=0$ only.}
	\label{fig:wrinkles_Re}
\end{figure}

We now extend the previous results to larger Reynolds numbers, up to 550. For these Reynolds numbers, the DNS data are available only in the small box $(8\pi, 3\pi)\delta$, so we must use a larger liquid viscosity, $\tilde{\nu_\ell} = 6~10^{-3}$, to reduce discretization errors; the results can however be  extrapolated to smaller viscosities, as we have seen that the wrinkles properties do not depend on $\tilde{\nu_\ell}$, at least in the case $Re_\delta = 100$ (see Fig.~\ref{fig:wrinkles_doppler}).

Results for the four characteristic wrinkle properties are plotted in Fig.~\ref{fig:wrinkles_Re} as a function of the Reynolds number for three values of the current $U_s/u^*$. The evolution of these quantities with $Re_\delta$ is similar to the case $U_s = 0$ already documented in Perrard {\it et al.}~\cite{Perrard2019}: The wrinkles tend to be more elongated in the wind direction (larger $\Lambda_x$ and smaller $\Lambda_y$) as $Re_\delta$ increases, the convection velocity $U_c/U_a$ rapidly falls off, and the wrinkle amplitude increases. Here again, the stronger effect of current is found for the streamwise length $\Lambda_x$, with a decrease with $U_s$ still compatible with Eq.~(\ref{eq:fitLx}) at larger $Re_\delta$; only the largest $Re_\delta=550$ deviates from the trend, which may originate from the limited computation time $T_\mathrm{max}$ (and hence stronger discretization effect) for this $Re_\delta$. For the other quantities, the variations with $Re_{\delta}$ do not show any significant dependence with $U_s$, thereby suggesting that the weak effects found at $Re_\delta=100$ can be extended to larger Reynolds numbers.

\begin{figure}[tb]
	\includegraphics[width=1\linewidth]{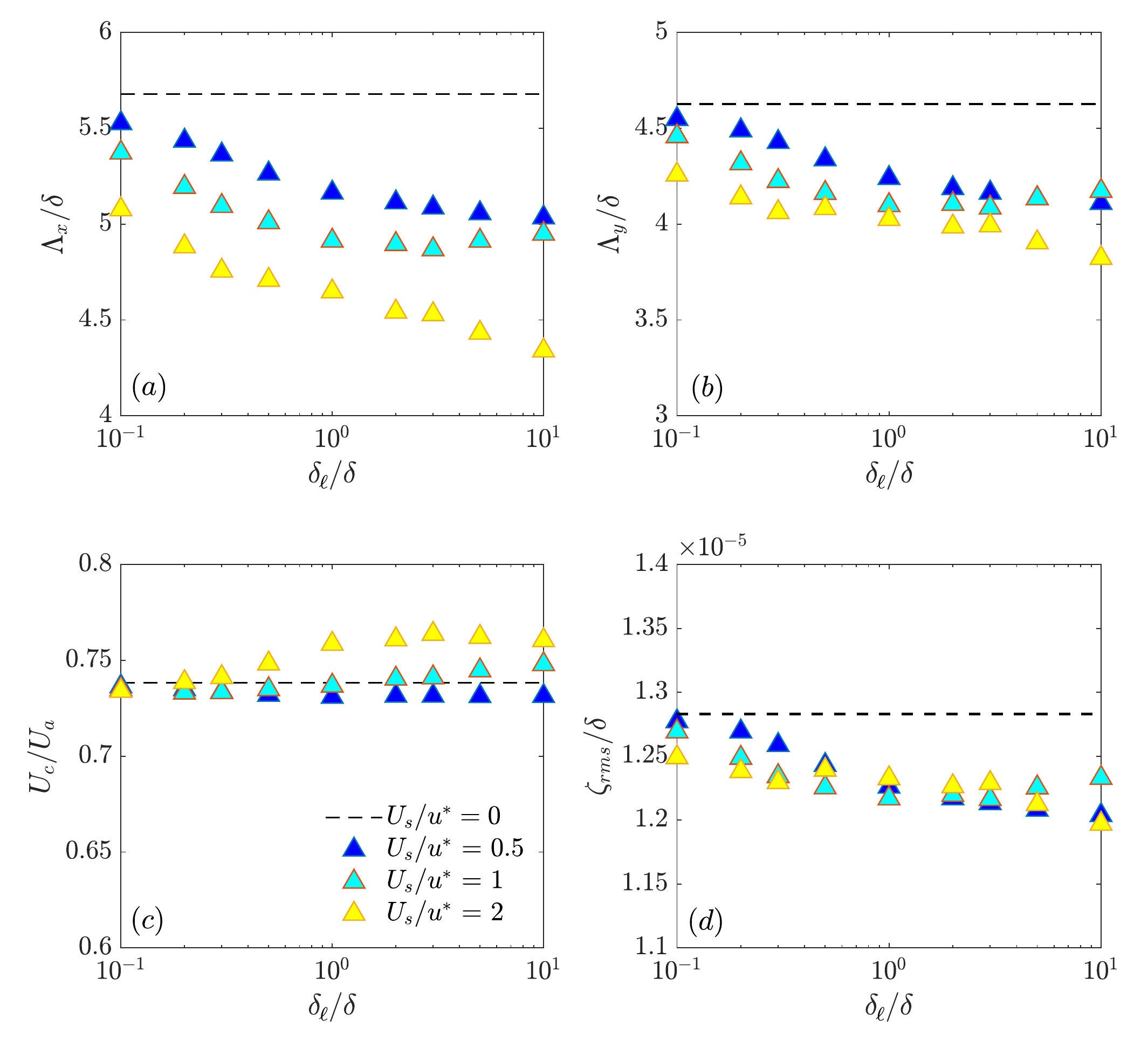}
	\caption{Wrinkle properties as a function of the normalized liquid layer thickness $\delta_\ell/\delta$ for various currents $U_s/u^*=0.5, 1, 2$. Reference values obtained without any current ($U_s=0$) are represented by the dashed black lines. Results are obtained at Re$_{\delta}$=100 for the large box $(60\pi,6\pi)\delta$, with a liquid viscosity of $\tilde{\nu_\ell}$ = 6~10$^{-3}$.}
	\label{fig:wrinkles_SJ}
\end{figure}

We finally consider the more realistic case of a sheared profile decreasing exponentially with depth [Eq.~(\ref{eq:exp_profile})], still in the direction of the wind ($\theta=0$). In addition to the normalized surface current $U_s/u^*$, we also consider now the influence of the normalized  liquid layer thickness $\delta_\ell/\delta$, restricting ourselves to the case $Re_\delta=100$ for which the data in the large box is available. The same four quantities characterizing the wrinkle properties are plotted as a function of the thickness ratio $\delta_\ell/\delta$ in Fig.~\ref{fig:wrinkles_SJ} for various surface velocities.  This ratio covers a wide range in practice: for wind-generated currents, laboratory experiments typically have $\delta_\ell \simeq 1$~cm in the liquid and $\delta \simeq 10$~cm in the air~\cite{caulliez2007turbulence}, yielding $\delta_\ell/\delta \simeq 0.1$; in the ocean, $\delta_\ell$ is typically 10~cm or more, while the boundary layer thickness $\delta$ can cover a wide range in unsteady conditions (as discussed in the introduction), yielding $\delta_\ell/\delta \ll 1$. For currents generated by other means, $\delta_\ell$ can be arbitrarily large, so the limit $\delta_\ell / \delta \gg 1$ is also relevant in general.

The results in Fig.~\ref{fig:wrinkles_SJ} show a slow variation of the wrinkle properties with $\delta_\ell/\delta$, bridging the reference case without drift as $\delta_\ell\rightarrow 0$ (dashed line) and the uniform current case as $\delta_\ell\rightarrow \infty$.  This confirms the filtering role of the liquid layer $\delta_\ell$ in the Doppler effect: the uniform current ($\delta_\ell\rightarrow \infty$) represents the bounding case with maximum effect, with a transition around $\delta_\ell / \delta \simeq O(1)$ towards no effect in the limit of a thin flowing liquid layer. We can conclude that the weak influence of uniform currents on the wrinkle properties is also valid for sheared currents, but with even weaker effects.

\section{Conclusion}
\label{sec:conclusion}

In this paper we investigated numerically the influence of a weak sheared current on the properties of the wind-generated wrinkles for a wind velocity below the onset and growth of regular waves. In that regime, the wrinkles are the statistically homogeneous and stationary response to the pressure fluctuations in the turbulent boundary layer, and their amplitude is governed by the viscosity of the liquid. We find that a longitudinal current tends to produce shorter and more fragmented wrinkles, whereas a transverse current simply tilts the wrinkles without modifying much their shape. In spite of these visual evidences, the overall effect of a longitudinal current remains weak: the energy-containing scale of the wrinkles only slightly decreases (about 5\% for the typical wind-generated surface current $U_s \simeq 0.6 u^*$ reported in the literature), and their amplitude is remarkably independent of the current. This confirms that the wrinkle properties described in Perrard {\it al.}~\cite{Perrard2019} are robust with respect to currents. 

This weak dependence of wrinkles on currents may have implications for the onset of regular waves at larger wind velocity. In Ref.~\cite{Perrard2019} we proposed that wrinkles form a base state from which regular waves are triggered, with a transition in friction velocity $u^*$ when the wrinkle amplitude $\zeta_\mathrm{rms}$ becomes of the order of the viscous sublayer thickness $\delta_\nu = \nu_a / u^*$, yielding a critical friction velocity for the onset of regular waves $u^*_c \simeq \nu_\ell^{1/5}$. According to this model, the feedback of the surface deformations on the turbulent boundary layer can no longer be neglected above this threshold, leading to a phase coherence between wind and waves, and hence a possible increase of energy transfers. Based on the observation made here regarding the independence of wrinkle amplitude from surface current, we may conclude that the critical friction velocity $u^*_c$  should be essentially independent of the current. However, the argument of Ref.~\cite{Perrard2019} is based on the wrinkle amplitude only, not on their shape, so an influence of the current on $u^*_c$ cannot be ruled out.  While the independence of $u^*_c$ with current is a reasonable assumption in the presence of a longitudinal current, for which the wrinkles remain aligned with wind, it is questionable for a transverse current: the cross-wind orientation of the wrinkles in that case probably induces stronger disturbances in the turbulent boundary layer, which could reduce the critical friction velocity $u^*_c$. Such a subtle dependence of the onset of regular waves in wrinkle geometry may contribute to the large variability of the critical velocities reported in the literature ($U_a \simeq 1-3$~m~s$^{-1}$, see Ref.~\cite{Paquier_2016}), with values usually larger in in controlled laboratory experiments than in open conditions where uncontrolled currents may be present.

\section*{Acknowledgements}
\label{sec:acknowledgements}

This work was supported by the  project ``ViscousWindWaves'' (ANR-18-CE30-0003) of the French National Research Agency, and by the project ``OVA'' of the LabeX LaSIPS (ANR-10-LABX-0040-LaSIPS) managed by the French National Research Agency under the "Investissements d'avenir" program (ANR-11-IDEX-0003-02). A.L.D. acknowledges the support from the Office of Naval Research under Grant \#N000141712310.

\bibliographystyle{unsrt}
\bibliography{biblio_WbyW}

\end{document}